\documentclass[fleqn,usenatbib]{mnras}

\usepackage{newtxtext,newtxmath}

\usepackage[T1]{fontenc}
\usepackage{ae,aecompl}
\usepackage{rotating}

\usepackage{longtable}
\usepackage{lscape}
\usepackage{balance}

\usepackage{graphicx}	
\usepackage{amsmath}	

\newcommand{\vsini}{$v$\,sin\,$i$}

\usepackage{enumerate}
\usepackage[british]{babel}

\title[]{Masses and compositions of three small planets orbiting the nearby M dwarf L231-32 (TOI-270) and the M dwarf radius valley}
\vspace{-1.5cm}
\author[\vspace{-0.5cm}V. Van Eylen et al.]{V. Van Eylen$^{1,2}$\thanks{E-mail: \textbf{v.vaneylen@ucl.ac.uk}},
N.\,Astudillo-Defru$^{3}$, 
X.\,Bonfils$^{4}$,
J.\,Livingston$^{5}$, 
T.\,Hirano$^{6}$,\newauthor
R.\,Luque$^{7,8}$,
K.\,W.\,F.\,Lam$^{9}$,
A.\,B.\,Justesen$^{10}$,
J.\,N.\,Winn$^{2}$,
D.\,Gandolfi$^{11}$,
G.\,Nowak$^{7,8}$,\newauthor
E.\,Palle$^{7,8}$,
S.\,Albrecht$^{10}$,
F.\,Dai$^{12}$,
B.\,Campos Estrada$^{13}$,
J.\,E.\,Owen$^{13}$,
D.\,Foreman-Mackey$^{14}$,\newauthor
M.\,Fridlund$^{15,16}$,
J.\,Korth$^{17}$, 
S.\,Mathur$^{7,8}$, 
T. Forveille$^{4}$,
T. Mikal-Evans$^{18}$,
H.~L.~M.~Osborne$^{1}$,\newauthor
C.~S.~K.~Ho$^{1}$,
J. M. Almenara$^{4}$, 
E. Artigau$^{19}$, 
O. Barrag\'an$^{20}$, 
S.~C.C.~Barros$^{21,22}$, 
F. Bouchy$^{23}$,\newauthor  
J. Cabrera$^{24}$, 
D. A. Caldwell$^{25}$, 
D. Charbonneau$^{26}$, 
P. Chaturvedi$^{27}$, 
W. D. Cochran$^{28}$, \newauthor
S. Csizmadia$^{24}$,
M. Damasso$^{29}$, 
X. Delfosse$^{4}$, 
J. R. De Medeiros$^{30}$, 
R. F. D\'iaz$^{31}$, 
R. Doyon$^{19}$,\newauthor 
M. Esposito$^{27}$, 
G. F\H ur\'esz$^{32}$, 
P. Figueira$^{33,21}$, 
I. Georgieva$^{15}$,
E. Goffo$^{11}$,
S. Grziwa$^{18}$,\newauthor %
E.\,Guenther$^{27}$, 
A. P. Hatzes$^{27}$, 
J. M. Jenkins$^{34}$, 
P. Kabath$^{35}$, 
E. Knudstrup$^{10}$, 
D. W. Latham$^{26}$, \newauthor
B. Lavie$^{23}$, 
C. Lovis$^{23}$, 
R.E. Mennickent$^{36}$, 
S. E. Mullally$^{37}$, 
F. Murgas$^{7.8}$,
N. Narita$^{38,39,40,7}$,\newauthor 
F. A. Pepe,$^{23}$ 
C. M. Persson$^{15}$,
S. Redfield$^{41}$, 
G. R. Ricker$^{18}$, 
N.~C.~Santos$^{21,22}$,\newauthor 
S.~Seager$^{18,42,43}$, 
L. M. Serrano$^{11}$,
A. M. S. Smith$^{24}$, 
A. Su\'{a}rez Mascare\~{n}o$^{8}$, 
J. Subjak$^{34}$,\newauthor 
J. D. Twicken$^{25,34}$, 
S. Udry$^{23}$, 
R. Vanderspek$^{42}$, and 
M. R. Zapatero Osorio$^{44}$\\ 
{~}\\
$^{1}$Mullard Space Science Laboratory, University College London, Holmbury St Mary, Dorking, Surrey RH5 6NT, UK\\
$^{2}$Department of Astrophysical Sciences, Princeton University, 4 Ivy Lane, Princeton, NJ, 08544, USA\\
$^{3}$Departamento de Matem\'atica y F\'isica Aplicadas, Universidad Cat\'olica de la Sant\'isima Concepci\'on, Alonso de Rivera 2850, Concepci\'on, Chile\\
$^{4}$Univ. Grenoble Alpes, CNRS, IPAG, F-38000 Grenoble, France\\
$^{5}$Department of Astronomy, Graduate School of Science, The University of Tokyo, Hongo 7-3-1, Bunkyo-ku, Tokyo, 113-0033, Japan\\
$^{6}$Department of Earth and Planetary Sciences, Tokyo Institute of Technology, 2-12-1 Ookayama, Meguro-ku, Tokyo 152-8551, Japan\\
$^{7}$Departamento de Astrof\'isica, Universidad de La Laguna, E-38206, Tenerife, Spain\\
$^{8}$Instituto de Astrof\'isica de Canarias, C/ V\'ia L\'actea s/n, E-38205, La Laguna, Tenerife, Spain\\
$^{9}$Zentrum f\"ur Astronomie und Astrophysik, Technische Universit\"at Berlin, Hardenbergstr. 36, 10623 Berlin, Germany\\
$^{10}$Stellar Astrophysics Centre, Deparment of Physics and Astronomy, Aarhus University, Ny Munkegade 120, DK-8000 Aarhus C, Denmark\\
$^{11}$Dipartimento di Fisica, Universit\`a degli Studi di Torino, via Pietro Giuria 1, I-10125, Torino, Italy\\
$^{12}$Division of Geological and Planetary Sciences, 1200 E. California Boulevard, Pasadena, CA, 91125, USA\\
$^{13}$Astrophysics Group, Imperial College London, Blackett Laboratory, Prince Consort Road, London SW7 2AZ, UK\\
$^{14}$Center for Computational Astrophysics, Flatiron Institute, 162 Fifth Avenue, New York, NY 10010, USA\\
$^{15}$Department of Space, Earth and Environment, Chalmers University of Technology, Onsala Space Observatory, 439 92 Onsala, Sweden\\
$^{16}$Leiden Observatory, Leiden University, postbus 9513, 2300RA Leiden, The Netherlands\\
$^{17}$Rheinisches Institut f\"ur Umweltforschung, Abteilung Planetenforschung an der Universit\"at zu K\"oln, Aachener Strasse 209, 50931 K\"oln, Germany\\
$^{18}$Department of Physics and Kavli Institute for Astrophysics and Space Research, Massachusetts Institute of Technology, Cambridge, MA 02139, USA\\
$^{19}$Institut de Recherche sur les Exoplan\`etes (IREx), D\'epartement de Physique, Universit\'e de Montr\'eal, C.P. 6128, Succ. Centre-Ville, Montr\'eal, QC, H3C 3J7, Canada\\
$^{20}$Sub-department of Astrophysics, Department of Physics, University of Oxford, Oxford, OX1 3RH, UK\\
$^{21}$Instituto de Astrof\'{i}sica e Ci\^{e}ncias do Espaço, Universidade do Porto, CAUP, Rua das Estrelas, 4150-762 Porto, Portugal\\
$^{22}$Departamento de F\'isica e Astronomia, Faculdade de Ci\^encias, Universidade do Porto, Rua do Campo Alegre, 4169-007 Porto, Portugal\\
$^{23}$Observatoire astronomique de l'Universit\'e de Gen\`eve, 51 ch des Maillettes, 1290 Versoix, Switzerland\\
$^{24}$Institut für Planetenforschung, Deutsches Zentrum f\"ur Luft- und Raumfahrt (DLR), Rutherfordstr. 2, 12489 Berlin, Germany\\
$^{25}$SETI Institute, 189 N. Bernardo Ave, Mt. View, CA 94043, USA\\
$^{26}$Center for Astrophysics | Harvard \& Smithsonian, 60 Garden Street, Cambridge MA 02138 USA\\
$^{27}$Th\"uringer Landessternwarte Tautenburg, Sternwarte 5, D-07778 Tautenberg, Germany\\
$^{28}$Center for Planetary Systems Habitability and McDonald Observatory, The University of Texas at Austin, Austin, TX 78730, USA\\
$^{29}$INAF -- Osservatorio Astrofisico di Torino, Via Osservaorio 20, I-10025 Pino Torinese, Italy\\
$^{30}$Departamento de F\'isica Te\'orica e Experimental, Universidade Federal do Rio Grande do Norte, Campus Universit\'ario, Natal, RN, 59072-970, Brazil \\
(affiliations continued after acknowledgments)\\
~\\
\vspace{-1.2cm}
}
\begin{document}
\label{firstpage}
\pagerange{\pageref{firstpage}--\pageref{lastpage}}
\maketitle

\clearpage
\begin{abstract}
We report on precise Doppler measurements of L231-32 (TOI-270), a nearby M dwarf ($d=22$\,pc, $M_\star = 0.39$~M$_\odot$, $R_\star = 0.38$~R$_\odot$), which hosts three transiting
planets that were recently discovered using data from the Transiting Exoplanet Survey Satellite (TESS). The three planets are 1.2, 2.4, and 2.1 times the size of Earth and have orbital periods of 3.4, 5.7, and 11.4 days. We obtained 29 high-resolution optical spectra with the newly commissioned Echelle Spectrograph for Rocky Exoplanet and Stable Spectroscopic Observations (ESPRESSO) and 58 spectra using the High Accuracy Radial velocity Planet Searcher (HARPS). From these observations, we find the masses of the planets to be $1.58 \pm 0.26$, $6.15 \pm 0.37$, and $4.78 \pm 0.43$~M$_\oplus$, respectively. The combination of radius and mass measurements suggests that the innermost planet has a rocky composition similar to that of Earth, while the outer two planets have lower densities. Thus, the inner planet and the outer planets are on opposite sides of the `radius valley' --- a region in the radius-period diagram with relatively few members, which has been interpreted as a consequence of atmospheric photo-evaporation. We place these findings into the context of other small close-in planets orbiting M dwarf stars, and use support vector machines to determine the location and slope of the M dwarf ($T_\mathrm{eff} < 4000$~K) radius valley as a function of orbital period.
We compare the location of the M dwarf radius valley to the radius valley observed for FGK stars, and find that its location is a good match to photo-evaporation and core-powered mass loss models. Finally, we show that planets below the M dwarf radius valley have compositions consistent with stripped rocky cores, whereas most planets above have a lower density consistent with the presence of a H-He atmosphere.
\end{abstract}



\section{Introduction}

The small, Earth-sized planets that are being discovered around other stars may or may not resemble our own Earth in terms of composition, formation history, and atmospheric properties. They are also a challenge to study, because they produce such small transit and radial-velocity (RV) signals. Fortunately, recent progress has been made on both of these fronts. 
The Transiting Exoplanet Survey Satellite (\textit{TESS}) was launched in April 2018 to conduct an all-sky survey and discover transiting planets around the nearest and brightest stars \citep{ricker2014}. Because the transit signal varies inversely as the stellar radius squared, searching small M~dwarf stars is of particular interest because there is a greater opportunity to find small planets. For the same reason, planets around M~dwarfs are
valuable targets for atmospheric studies through transmission spectroscopy.
To understand which planets have a rocky composition and whether they are likely to have atmospheres, precise radius measurements from transit surveys need to
be paired with mass measurements from dynamical observations such as precise RV measurements. To this end, the novel Echelle Spectrograph for Rocky Exoplanet and Stable Spectroscopic Observations \citep[ESPRESSO,][]{pepe2010,pepe2014,pepe2020} at the Very Large Telescope (VLT) provides an unprecedented RV precision.

Here we present the result of an ESPRESSO campaign (ESO observing program 0102.C-0456) to characterise three small (1.1, 2.3, and 2.0 $R_\oplus$) transiting planets around L231-32 (TOI-270), a nearby (22~pc), bright ($K = 8.25$, $V = 12.6$), M3V dwarf star ($M_\star = 0.39$~M$_\odot$, $R_\star = 0.38$~R$_\odot$), as well as four additional ESPRESSO observations obtained as part of observing programs 1102.C-0744 and 1102.C-0958. We also used data from the High Accuracy Radial velocity Planet Searcher (HARPS) program for M-dwarf planets amenable to detailed atmospheric characterisation (ESO observing program 1102.C-0339). These planets were observed to transit by \textit{TESS} in three subsequent campaigns, each lasting about 27 days. The transit signals were described and validated by \cite{guenther2019}. Because of the characteristics of the host star and its planets, L231-32 is a prime target for exoplanet atmosphere studies, which are ongoing with the Hubble Space Telescope \citep[HST, program id GO-15814, PI Mikal-Evans;][]{mikalevans2019}. Furthermore, simulations have shown these planets are highly suitable for atmospheric characterisation with the James Webb Space Telescope \citep[JWST;][]{chouqar2020}. Further transit observations with e.g.\ \textit{TESS} or other space telescopes may reveal transit timing variations (TTVs) which can be used to constrain the planet masses independently of RVs.

Our mass measurements for these three planets allow us to constrain their possible compositions. The planets are located on both sides of the radius valley, which separates close-in super-Earth planets from sub-Neptune planets  \citep[e.g.][]{owen2013,lopez2013,fulton2017,vaneylen2018}, and we show how their compositions can be interpreted in this context. We furthermore compare the properties of L231-32's planets with those of other small planets with precisely measured masses, radii, and periods, and use this sample to measure the location of the M dwarf radius valley and its slope as a function of orbital period. 

This paper is organised as follows. In Section~\ref{sec:observations}, we describe the \textit{TESS} transit observations, and ESPRESSO and HARPS RV observations. In Section~\ref{sec:star}, we derive the parameters of the host star, by combining high resolution spectra with other sources of ancillary information. In Section~\ref{sec:planet}, we describe the approach to modeling L231-32 and the properties of its planets. In Section~\ref{sec:results_toi270}, we show the resulting properties of L231-32's planets and discuss the composition of the planets. In Section~\ref{sec:radiusvalley}, we compare their properties to radius valley predictions and to other planets orbiting M dwarf stars, and measure the location and slope of the M dwarf radius valley. Finally, in Section~\ref{sec:conclusions}, we provide a brief summary and conclusions.

\section{Observations and modeling}
\label{sec:observations}

\subsection{TESS photometry}
\label{sec:tessphotometry}

L231-32 (TOI-270; TIC~259377017) was observed by the \textit{TESS} mission \citep{ricker2014} during three 27-day sectors, namely sectors 3, 4, and 5, between 20 September 2018 and 11 December 2018. It was observed on CCD 4 of camera 3 in sector 3 and 4, and on CCD 3 of camera 3 in sector 5. The star was pre-selected \citep{stassun2018} and was observed in a 2-minute cadence for the whole duration of these sectors. During the TESS extended mission, the target was re-observed in sectors 30 (between 23 September 2020 and 19 October 2020) and 32 (between 20 November 2020 and 16 December 2020) in the 2-minute cadence mode. The data observed in sector 30 was taken on CCD 4 of camera 3, while the data observed in sector 32 was taken on CCD 3 of camera 3. The data were reduced by the \textit{TESS} data processing pipeline developed by the Science Processing Operations Center \citep[SPOC,][]{jenkins2016}, and the transits of three planet candidates were detected in the SPOC pipeline and promoted to TESS object of interest (TOI) status by the TESS science team.

We started radial velocity (RV) observations to confirm and measure the mass of these transiting planets with ESPRESSO and with HARPS, on 9 February and 1 January 2019, respectively (see Section \ref{sec:observations_spectro}). As these observations were ongoing, \cite{guenther2019} also reported on the validation of these planets, by performing a statistical analysis of the \textit{TESS} observations, as well as obtaining ground-based seeing-limited photometry coordinated through the \textit{TESS} Follow-up Observing Program (TFOP)\footnote{\url{https://tess.mit.edu/followup/}}.

We downloaded the \textit{TESS} photometry from the Mikulski Archive for Space Telescopes (MAST\footnote{\url{https://archive.stsci.edu/tess}}) and started our analysis using the presearch data conditioning (PDC) light curve reduced by SPOC. We searched for additional transit signals using a Box Least-Square (BLS) algorithm \citep{kovacs2002} and the `D\'etection Sp\'ecialis\'ee de Transits' (DST) algorithm \citep{cabrera2012} for additional transit signals but found no evidence for any transiting planets in addition to the three that were alerted. In Section~\ref{sec:transitmodel}, we describe our approach to the modeling of the \textit{TESS} photometry.

\subsection{Spectroscopic observations}
\label{sec:observations_spectro}

\subsubsection{ESPRESSO} 
\label{sec:rv_espresso}
We obtained 26 high-resolution spectroscopic observations of L231-32 between 9 February and 22 March 2019 using ESPRESSO \citep[][]{pepe2014,pepe2020} on the 8.2~m Very Large Telescope (VLT; Paranal, Chile) as part of observing program 0102.C-0456. Four additional observations were obtained as part of observing programs 1102.C-0744 and 1102.C-0958. ESPRESSO is a relatively novel instrument at the VLT which was first offered to the community in October 2018.

Each observation has an integration time of 1200~sec, a median resolving power of 140,000, and a wavelength range of 380-788 nm. We used the slow read-out mode, which uses a 2x1 spatial by spectral binning. Observations were taken in high-resolution (HR) mode. One of the observations was flagged as unreliable due to a detector restart just before the exposure, and we excluded this data point from the analysis as a restart can result in additional noise. This leaves a total of 29 ESPRESSO observations that we use in our subsequent analysis.
 
To determine a wavelength-calibration solution, daytime ThAr measurements were taken and the source was observed with simultaneous Fabry-P\'erot (FP) exposures, following the procedure outlined in the ESPRESSO user manual\footnote{\url{https://www.eso.org/sci/facilities/paranal/instruments/espresso/ESPRESSO_User_Manual_P102.pdf}}. The signal-to-noise ratio (SNR) for individual spectra at orders 104 and 105, which are both centered at 557 nm, ranges from 9 to 40, with a median SNR of 29.

To calibrate and reduce the data we used the publicly available pipeline for ESPRESSO data reduction\footnote{\url{http://eso.org/sci/software/pipelines/espresso/espresso-pipe-recipes.html, version 2.0}}, together with the ESO Reflex tool \citep{freudling2013}. This tool uses the science spectra and all associated calibration files (i.e.\ bias frames, dark frames, led frames, order definitions, flat frames, FP wavelength calibration frame, Thorium-FP calibration, FP-Thorium calibration, fiber-to-fiber efficiency exposure, and a spectrophotometric standard star exposure) to reduce and calibrate the raw spectra, and provide 1-dimensional and 2-dimensional reduced spectra\footnote{See the ESPRESSO pipeline user manual for details, \url{ftp://ftp.eso.org/pub/dfs/pipelines/instruments/espresso/espdr-pipeline-manual-1.2.3.pdf}}. All required calibration frames are automatically associated with the raw science frames and were simultaneously downloaded from the ESO archive\footnote{\url{http://archive.eso.org/}}. We followed the standard pipeline routines for the ESPRESSO data reduction pipeline using ESO Reflex. 
The radial velocities were computed following the procedure described in \cite{Astudillo-Defru2017b}. A slight adaptation to ESPRESSO data was introduced to construct the stellar template by combining the two echelle orders covering a common spectral range. The template is then Doppler shifted and we maximised its likelihood with each 2-dimensional reduced spectrum.
The mean RV precision is 0.47~m~s$^{-1}$. The resulting RV observations are listed in Table~\ref{tab:rvdata}. 

\subsubsection{HARPS}

We used HARPS \citep[][]{mayor2003} on the the La Silla 3.6m telescope to gather 58 additional spectra (program id.\ 1102.C-0339). These high-resolution spectra, with a resolving power of 115,000, were obtained between 1 January 2019 and 17 April 2019, spanning 89 days. We fixed the exposure time to 1800 s, resulting in a total equivalent to 29 h of open shutter time. The readout speed was set to 104 kHz. To prevent possible contamination from the calibration lamp in the bluer zone of the spectral range, we elected to put the calibration fiber on the sky.

Raw data were reduced with the dedicated HARPS Data Reduction Software \citep{LovisPepe2007}. The resulting spectra have a signal-to-noise ratio ranging between 14 and 29 at 550 nm, with a median of 22. 

We then extracted RVs, again following \cite{Astudillo-Defru2017b}, resulting in an RV extraction consistent with how RVs were extracted for ESPRESSO (see Section~\ref{sec:rv_espresso}). The mean RV precision of the HARPS observations is 2.05~m~s$^{-1}$ and the data have a dispersion of 5.10~m~s$^{-1}$. The resulting RVs and stellar activity indices are listed in Table~\ref{tab:rvdata_harps}.
The mean RV precision of the HARPS observations is 2.17~m~s$^{-1}$. 
The HARPS timestamps were converted to Barycentric Dynamical
Time (BJD$_\mathrm{TDB}$) for consistency with ESPRESSO and \textit{TESS} observations.
The resulting RVs and stellar activity indices are listed in Table~\ref{tab:rvdata_harps}.

\section{Stellar parameters}
\label{sec:star}	

\subsection{Fundamental stellar parameters}

We co-added the ESPRESSO spectra and analysed the combined spectrum of L231-32 to estimate its spectroscopic parameters. We used {\tt SpecMatch-Emp} \citep{yee2017}, which is known to provide accurate estimates for late-type stars. Following the prescriptions described in \citet{hirano2018}, we lowered the spectral resolution of the combined ESPRESSO spectrum from $140,000$ to $60,000$ and stored the spectrum in the same format as Keck/HIRES spectra before inputting it into {\tt SpecMatch-Emp}. 

Using {\tt SpecMatch-Emp}, we determined the stellar effective temperature ($T_\mathrm{eff,sm}$), stellar radius ($R_{\star,\mathrm{sm}}$), and metallicity ($\mathrm{[Fe/H]_{sm}}$), and found $T_\mathrm{eff,sm} = 3506 \pm 70$~K, $R_{\star,\mathrm{sm}} = 0.410 \pm 0.041~R_\odot$, and $\mathrm{[Fe/H]_{sm}} = -0.20 \pm 0.12$. 
To obtain precise stellar parameters, we combined the spectroscopic information with a distance measurement of the star based on the \textit{Gaia} parallax \citep[$44.457 \pm 0.027$~mas,][]{gaia2018} and the apparent magnitude of the star from the 2MASS catalogue 
\citep[$m_\mathrm{K_s,2MASS} = 8.251 \pm 0.029$,][]{skrutskie2006}.
For the Gaia observations, we include an additional uncertainty as reported by \cite{stassun2018}, who find a systematic error of $0.082 \pm 0.033$~mas. We adopted a conservative systematic error of $0.115$~mas and add this in quadrature to the internal error on the parallax measurement of L231-32. This results in a distance estimate of $d_\mathrm{Gaia} = 22.453 \pm 0.060$~pc. 

To determine stellar parameters that combine the information from the apparent magnitude, distance, and spectra, we implemented a Markov Chain Monte Carlo (MCMC) code to estimate the final stellar parameters. For this, we defined the log likelihood ($\log L$) as a function of the stellar radius ($R_{\star}$) and apparent magnitude ($m_\mathrm{K_{s}}$) as 
\begin{equation}
\log L \propto \frac{(R_{\star}-R_{\star,\mathrm{sm}})^2}{\sigma_{R_{\star,\mathrm{sm}}}^2}
+ \frac{(m_{K_s}-m_{K_\mathrm{s,2MASS}})^2}{\sigma_{m_{K_\mathrm{s,2MASS}}}^2}.
\label{eq:chisq}
\end{equation}
The parameters $R_{\star}$ and $m_\mathrm{K_{s}}$ are related to each other through the empirical relations determined by \cite{mann2015}. These relations show how $R_\star$ depends on [Fe/H] and the absolute magnitude $M_\mathrm{K_s}$, which in turn is related to the apparent magnitude and the distance through $m_\mathrm{K_s} - M_\mathrm{K_s} = 5 \log d - 5$. We imposed Gaussian priors on [Fe/H] and $d$, i.e.\

\begin{equation}
p_\mathrm{prior} \propto \exp\left(-\frac{(\mathrm{[Fe/H]}-\mathrm{[Fe/H]}_\mathrm{sm})^2}{2\sigma_{\mathrm{[Fe/H]_{sm}}}^2}
- \frac{(d-d_\mathrm{Gaia})^2}{2\sigma_{d_{Gaia}}^2}\right). 
\label{eq:prior}
\end{equation}

For $m_{K_s}$ and $R_\star$, we used a uniform prior distribution. We then sampled the likelihood and prior from Equations~\ref{eq:chisq} and \ref{eq:prior} using a customized MCMC implementation \citep{hirano2016}, which employs the Metropolis-Hastings algorithm and automatically optimises the chain step sizes of the proposal Gaussian samples so that the total acceptance ratio becomes $20-30\%$ after running $\sim10^6$ steps. From the MCMC posterior sample we determined the median and $15.87$ and $84.13$ percentiles to report best values and their uncertainties, which are shown in Table~\ref{tab:parameters}. We further determined the stellar mass ($M_\star$) from its corresponding empirical relation based on $m_{K_s}$ \citep[][Equation 5]{mann2015}. For both stellar radius and stellar mass, we take into account the uncertainty of the empirical relationships, which are $2.7\%$ and $1.8\%$, respectively \citep{mann2015}. Since the effective temperature ($T_\mathrm{eff}$) does not affect the empirical relations, we adopt the spectroscopic effective temperature as our final value ($T_\mathrm{eff} = T_\mathrm{eff,sm}$). We use $T_\mathrm{eff}$ and $R_\star$ to determine the stellar luminosity ($L$), and finally, from mass and radius we also calculated the stellar density ($\rho_\star$) and surface gravity ($\log g$). Interstellar extinction was neglected, as the star is relatively nearby. All these values are reported in Table~\ref{tab:parameters}. 

These values can be compared with the stellar parameters derived by \cite{guenther2019}. For example, the stellar mass and radius determined here, $0.386 \pm 0.008~M_\odot$ and $0.378 \pm 0.011~R_\odot$, are consistent with the mass and radius determined by \cite{guenther2019}, i.e.\ $0.40 \pm 0.02~M_\odot$ and $0.38 \pm 0.02~R_\odot$, respectively. The values determined here make use of a high-resolution combined ESPRESSO spectrum in addition to distance and magnitude information and are slightly more precise. Similarly, the temperature determined here, i.e.\ $3506 \pm 70~K$ is consistent with the value determined by \cite{guenther2019}, i.e. $3386^{+137}_{-131}~K$, and more precise.

\subsection{Stellar rotation}
\label{sec:rotation}

We also investigated the activity indicators from the spectra (see Section~\ref{sec:observations_spectro}) to estimate the stellar rotation period. In doing so, we focused on the HARPS observations, which span a longer baseline than the ESPRESSO observations and which are therefore more suitable to determine the stellar rotation period. We computed the Generalised Lomb-Scargle periodogram \citep[GLS;][]{zechmeister2009} of both $H_\alpha$ and Na~D and found consistent periods of 
$P=54.0 \pm 2.4$~d and 
$P=61.5 \pm 4.0$~d, 
respectively.
We finally also looked at the full-width at half maximum (FWHM) of the cross correlation function (CCF), which was extracted for HARPS observations directly by the Data Reduction Software\footnote{\url{http://www.eso.org/sci/facilities/lasilla/instruments/harps/doc/DRS.pdf}}, and find $P=57.5 \pm 5.7$~days.
We also searched the \textit{TESS} light curve for rotational modulation. The PDC pipeline \citep{Stumpe2012,smith2012,Stumpe2014} produces high quality light curves well-suited for transit searches. However, stellar rotation signals can be removed by the PDC photometry pipeline, so we used the {\tt lightkurve} package \citep{lightkurve} to produce systematics-corrected light curves with intact stellar variability. {\tt lightkurve} implements pixel-level decorrelation \citep[PLD;][]{Deming2015} to account for systematic noise induced by intra-pixel detector gain variations and pointing jitter. We normalized and concatenated the PLD-corrected light curves, then computed a GLS periodogram of the full time series. A sine-like signal is clearly visible, and GLS detects a significant $\sim$1.1 ppt signal at $57.90 \pm 0.23$ days. We also analysed the full time series with a pipeline that combines three different methods (a time-period analysis based on wavelets, auto-correlation function, and composite spectrum) and that has been applied to tens of thousand of stars \citep[e.g.][]{2014A&A...572A..34G, 2017A&A...605A.111C,2019FrASS...6...46M, 2019ApJS..244...21S}. The time-period and composite spectrum analyses find a signal around 46-49 days. The auto-correlation function does not converge due to the too short length of the observations, not allowing us to confirm the rotation period with this pipeline. Given the length of the data, it can still be possible that we measure a harmonic of the real rotation period. Since the total time series only spans $3 \times 27$~days, it is difficult to ascertain the veracity of this signal, but it appears consistent with the values determined from the RV activity indicators. For completeness, we also searched for signals in the sector-combined PDC light curve, but detected only short-timescale variability. 

We explored the possibility of estimating the stellar rotation period from its relationship with the $R^\prime_{HK}$ \citep[e.g.,][]{Astudillo2017a}. However, near the Ca II H\&K lines, the HARPS data set has an extremely low flux (with a median SNR of about 0.6) that limits the precision of this approach to measure the stellar rotation rate. We find $\log(R^\prime_{HK})=-5.480 \pm0.238$, and estimate a rotation period of $88 \pm 32$~days from this approach.

Based on the combination of RV stellar activity indicators, and the \textit{TESS} photometry, it appears likely that the stellar rotation period is approximately 58 days, which is consistent with typically observed rotation periods for M dwarf stars in this mass range, i.e.\ $\approx 20-60$~days \citep{newton2016}.

\section{Orbital and planetary parameters}
\label{sec:planet}

We modeled the \textit{TESS} light curve and ESPRESSO RV data using the publicly available {\tt exoplanet} code \citep{exoplanet:exoplanet}. This tool can model both transit and RV observations using a Hamiltonian Monte Carlo (HMC) scheme implemented in Python in {\tt pymc3} \citep{exoplanet:pymc3}, and has been used to model photometric and spectroscopic observations of other exoplanets \citep[e.g.][]{plavchan2020,kanodia2020,stefansson2020}. Below we describe the ingredients of our model and the procedure for optimising and sampling the orbital planetary parameters.

\subsection{Light curve model}
\label{sec:transitmodel}

\subsubsection{Transit model}
\label{sec:transitphysicalmodel}

We used {\tt exoplanet} to model the \textit{TESS} transit light curve, which makes use of {\tt starry} \citep{exoplanet:luger18,Agol2020} to calculate planetary transits. {\tt Starry} implements numerically stable analytic planet transit models with polynomial limb darkening---a generalisation of \citet{mandel2002}---along with their gradients. 
The transit model contains seven parameters for each planet ($i \in \{b,c,d\}$), i.e.\ the orbital period ($P_i$) and transit reference time ($T_{0,i}$), the planet-to-star ratio of radii ($R_{p,i}/R_\star$), 
the scaled orbital distance ($a_i/R_\star$), 
the impact parameter ($b_i$), and the eccentricity ($e_i$) and argument of periastron ($\omega_i$); furthermore, there are two stellar limb darkening parameters ($q_1$ and $q_2$) which are joint for all three planets. 
We use a uniform prior for $P_i$ and $T_{0,i}$ centered on an initial fit of the planet signals, with a broad width of $0.01$ and $0.05$ days, respectively, which encompasses the final values. We sample the ratio of radii uniformly in logarithmic space. For $b_i$ we sample uniformly between $0$ and $1$. 
We do not directly input a prior distribution for $a_i/R_\star$, because this parameter is directly constrained by $\rho_\star$ and the other transit parameters. Instead, we input $M_\star$ and $R_\star$ to the model using a normal distribution with mean and sigma as determined in Section~\ref{sec:star}.
The eccentricity of systems with multiple transiting planets is low but not necessarily zero \citep{vaneylen2015,xie2016,vaneylen2019}. We therefore do not fix eccentricity to zero, but place a prior on the orbital eccentricity, of a Beta distribution with $\alpha = 1.52$ and $\beta = 29$ \citep{vaneylen2019}. We sample $\omega$ uniformly between $-\pi$ and $\pi$. 
We adopt a quadratic limb darkening model with two parameters, which we reparametrize following \cite{exoplanet:kipping13} to facilitate efficient uninformative sampling.

\subsubsection{Gaussian Process noise model}
\label{sec:transitnoisemodel}

To model correlated noise in the \textit{TESS} light curve we adopt a Gaussian process model \citep{rasmussen2006,foremanmackey2017}. We adopt a stochastically driven damped harmonic oscillator (SHO) for which the power spectral density is defined as 
\begin{equation}
\label{eq:shokernel}
 S(\alpha) = \sqrt{\frac{2}{\pi}} \frac{S_0 \alpha_0^4}{(\alpha^2 - \alpha_0^2)^2 +     2 \alpha_0^2 \alpha^2 /Q_0^2 },
\end{equation}
where $\alpha_0$ is the frequency of the undamped oscillator and $S_0$ is proportional to the power at $\alpha = \alpha_0$. The SHO kernel is similar to the quasi-periodic kernel, which has been used extensively to model stellar activity \citep[e.g.][]{haywood2014,rajpaul2015,grunblatt2015}, but can be computed significantly faster and therefore facilitates a joint-fit of the transit observations as well as both HARPS and ESPRESSO RV data. Since we do not a priori know the values of $S_0$ and $\alpha_0$ we adopt a broad prior and relatively arbitrary starting value, in the form of a normal distribution, $\mathcal{N}(\mu,\sigma)$ where $\mu$ and $\sigma$ are the mean and standard deviation of the distribution. We adopt $\log \alpha_0 \sim \mathcal{N}(\log(2\pi/10), 10)$ and $\log S_0 \sim \mathcal{N}(\log(\sigma_{\mathrm{phot}}^2), 10)$, where $\sigma_{\mathrm{phot}}^2$ is the variance of the \textit{TESS} photometry. To limit the number of free parameters, we set $Q_0=1$.

Furthermore, we add a mean flux parameter ($\mu_\mathrm{norm}$) to our model. Since we normalized the light curve to zero, we place a broad prior of $\mu_\mathrm{norm} \sim \mathcal{N}(0,10)$. Finally, we include a noise term, which we fit as part of the Gaussian process model. We initalise this term based on the variance of the \textit{TESS} photometry, with a wide prior. All parameters and priors are summarised in Table~\ref{tab:priors}.

\subsection{Radial Velocity model}
\label{sec:rvmodel}

\subsubsection{Planet orbital model}
\label{sec:rvphysicalmodel}
We model the radial velocity (RV) variations of the host star using a Keplerian model for each planet, as implemented in {\tt exoplanet}. 
For each planet, we model the planet mass ($M_i$) which is associated to a RV semi-amplitude ($K_i$). The other parameters that determine the planet orbit ($P_i$, $T_{0,i}$, $e_i$, and $\omega_i$) were already defined in 
Section~\ref{sec:transitmodel}. 
For $M_i$ we adopt broad Gaussian priors
centered on initial guesses of the planet mass, made based on the observed amplitude of the RV curve.
Although some RV observations were taken during transit, we did not model the Rossiter-McLaughlin (RM) effect \citep{rossiter1924,mclaughlin1924}, as its RV amplitude is small relative to our RV precision.\footnote{Based on the estimated rotation period of 58 days (see Section~\ref{sec:rotation}) and the stellar radius, we find a stellar rotation speed, i.e.~maximum \vsini~for $i=90^\circ$, of $330$~m~s$^{-1}$. Given transit depths of the planets, impact parameters, and stellar limb darkening we estimate the maximum RM RV amplitude (assuming the planet's orbit is aligned with the rotation of the star), to be $\approx 0.5$~m~s$^{-1}$ \citep{albrecht2011}.}

\subsubsection{Noise model}
\label{sec:rvnoisemodel}

In addition to the formal uncertainty on each RV observation ($\sigma_\mathrm{rv}$, see Section~\ref{sec:observations_spectro} and Table~\ref{tab:rvdata}), we define an additional `jitter' noise term ($\sigma_\mathrm{2,rv}$) which is added in quadrature. We model this with a wide Gaussian prior as $\log \sigma_\mathrm{2,rv} \sim \mathcal{N}(\log(1),5)$.

We furthermore model the noise using a Gaussian process model, to model any correlation between the RV observations in a flexible way. This approach has been shown to reliably model stellar variability \citep[e.g.][]{haywood2014}. 
To do so, we again adopt the SHO kernel that was described in Section~\ref{sec:transitnoisemodel}. As the \textit{TESS} light curve is modified and filtered, we do not necessarily expect the photometric noise to occur in a similar way as noise in the RV observations, and so the two GP models are kept independent. 

The SHO kernel is defined as in Equation~\ref{eq:shokernel}, where we now have three hyperparameters $S_1$, $\alpha_1$, and $Q_1$. In all cases, we adopt broad priors. The hyperparameter $\alpha_1$ can be thought of as a periodic term. We therefore initialise it based on the expected stellar rotation, which we expect to be at around $55$~days (see Section~\ref{sec:rvselection}). The list of priors is shown in Table~\ref{tab:priors}.

\subsubsection{Comparing different RV models}
\label{sec:rvselection}

As outlined in Section~\ref{sec:rvnoisemodel}, we adopt a Gaussian process model to reliably estimate the planet masses from the RV observations, where the GP component is used to model the stellar rotation and activity in a flexible manner. We assessed whether this model is suitable for these observations.

We compared several possible RV models. In the first one, the RVs are modelled without taking into account any component describing stellar rotation (i.e. a `pure' 3-planet model, without GP). This model appears to perform significantly worse at modelling the RV observations than the models with a GP. 
Notably, it results in a fit for which the residuals have a distinctly correlated structure, and the jitter terms are significantly higher, suggesting a component is missing from the fit. This is not surprising, as we know the stellar rotation period of about 57 days (see Section~\ref{sec:rotation}) is likely to influence the RV signal. Even so, the resulting best-fit masses are fully consistent with the GP approach to better than $1\sigma$, providing confidence in our fitting approach and suggesting that L231-32 is a remarkably quiet M star. 

We also explored models in which we replaced the GP component with a polynomial trend instead, where we explored several different orders. In particular, a third order polynomial results in a fit where the polynomial resembles a sinusoid with a `peak to peak' period of around 50 days, visually similar to the GP model, suggesting it may similarly capture a quasi-periodic stellar rotation. Once again the planet masses are remarkably consistent, to better than $1\sigma$ for all three planets. Another model, in which we included a fourth `planet' (without polynomial trend), once again provides fully consistent masses; the orbital period of this `planet' was $\sim$63 days, consistent with the stellar rotation period found in Section~\ref{sec:rotation}.
We furthermore explored the specific choice of a GP kernel. For this, we used \texttt{RadVel}\footnote{\url{https://github.com/California-Planet-Search/radvel}} \citep{2018PASP..130d4504F}. With \texttt{RadVel}, we modeled the RV data using a quasi-periodic kernel, which is similar to the SHO kernel described in Section~\ref{sec:rvnoisemodel}, but the SHO kernel has properties that make it significantly faster to calculate (see Section~\ref{sec:rvnoisemodel}). This makes the SHO kernel more suitable for performing a joint transit-RV fit. Comparing the best-fit masses using both kernels, we find that they are consistent to a fraction of $\sigma$. 

\begin{figure*}
	\includegraphics[width=0.87\textwidth]{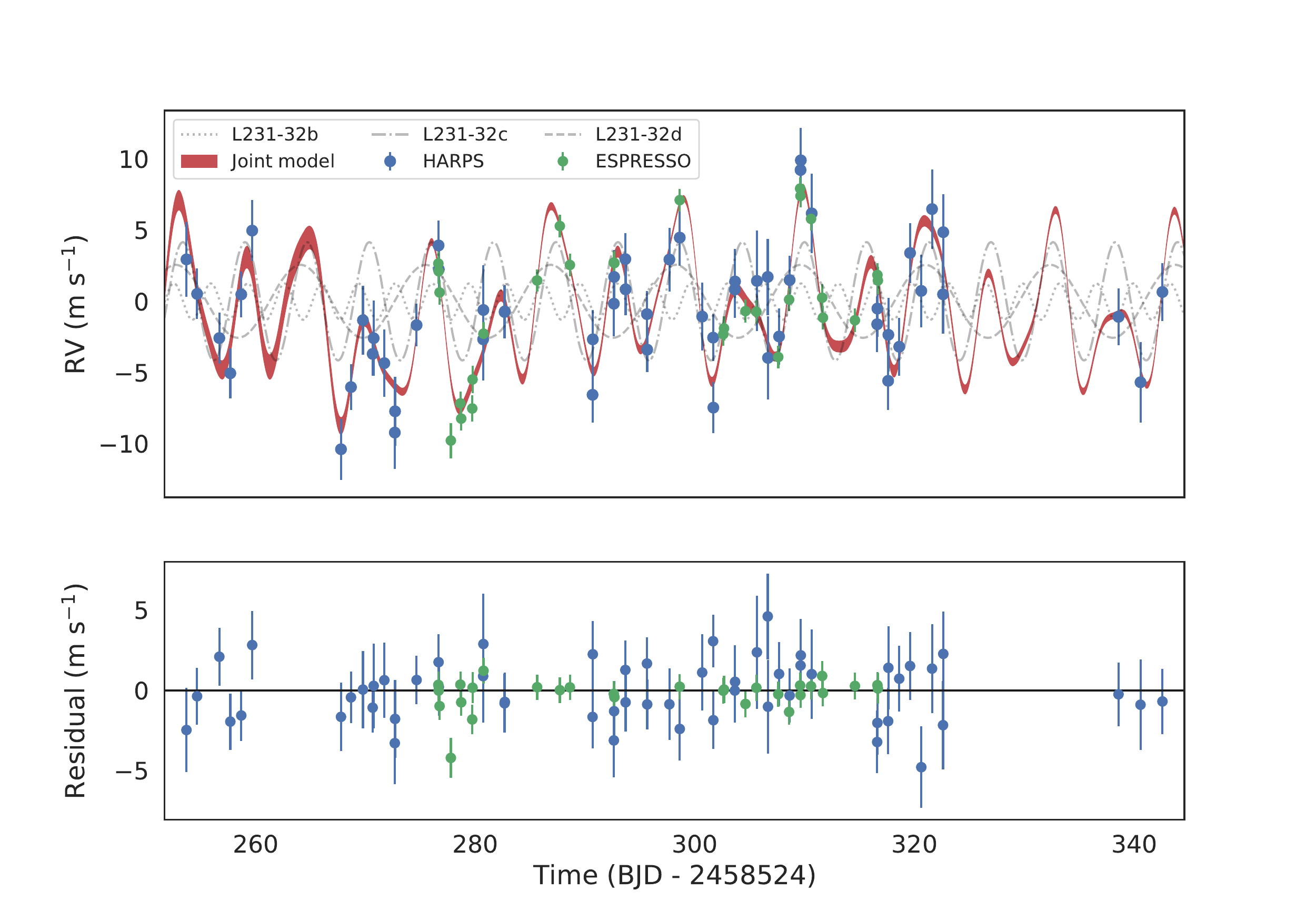}\\
	\caption{Top panel: ESPRESSO (green) and HARPS (blue) RV measurements and the best-fitting models for each of the three planets, and the joint model. The spread in the joint model represents the spread in GP parameters. The bottom panel shows the residuals to the joint model. The uncertainties represent the quadratic sum of the formal uncertainty and `jitter' uncertainty. The RMS of the residuals is 1.86~m~s$^{-1}$ for HARPS, and 0.95~m~s$^{-1}$ for ESPRESSO, respectively.}
	\label{fig:rv_quadratic}
\end{figure*}

In summary, these different model choices all result in very similar mass estimates, and the use of the kernel adopted here results in virtually the same results as using a quasi-periodic kernel. We therefore adopt the GP model with the SHO kernel described in Section~\ref{sec:rvnoisemodel}, as this can be calculated efficiently allowing for a joint fit with the transit data, and as a GP model can flexibly model the suspected stellar rotation signal in a reliable way \citep[e.g.][]{haywood2014}.
The resulting RV model is shown in Figure~\ref{fig:rv_quadratic}. We further calculated the root-mean square (RMS) of the residuals to the best fit. We find 1.86~m~s$^{-1}$ for HARPS, and 0.95~m~s$^{-1}$ for ESPRESSO. These small values confirm the quality of the fit and showcase the precision the ESPRESSO instrument is capable of.

\subsection{Joint analysis model}
\label{sec:joint}

\begin{figure*}
	\includegraphics[width=\textwidth]{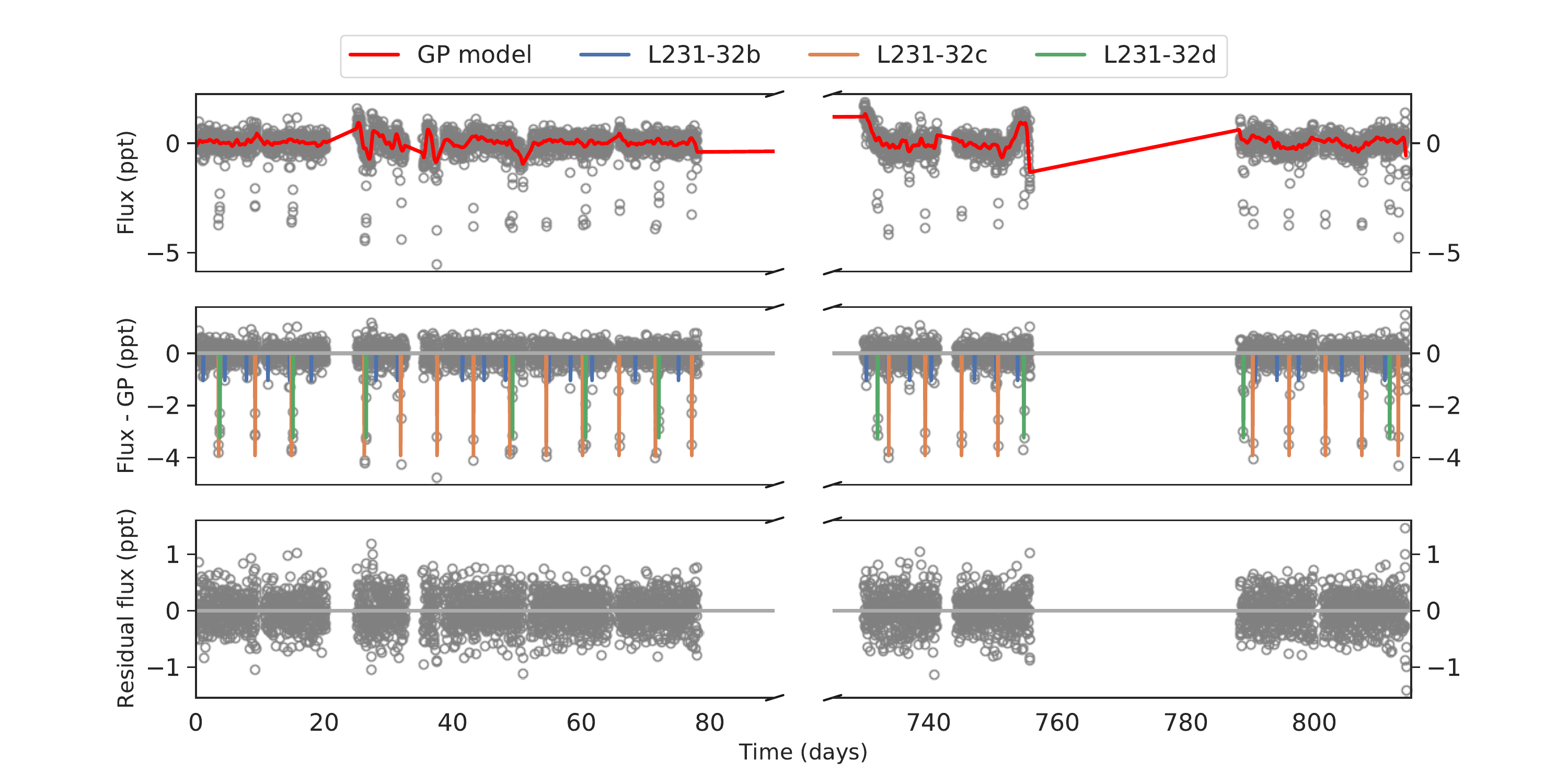}
	\caption{The \textit{TESS} light curve based on observations in sectors 3, 4, 5, and sectors 30 and 32 (data binned for clarity). The top panel shows the \textit{TESS} data with the best GP model, the middle panel shows the transit fits, and the bottom panel shows the residuals to both the GP and transit model. The three planets orbit near resonances, with planet $c$ and $b$ near a 5 to 3 resonance, and planets $d$ and $c$ near a 2 to 1 resonance.}
	\label{fig:timeseries}
\end{figure*}

We now combine the transit model and the RV model to run a joint fit of the \textit{TESS} and ESPRESSO/HARPS observations. To summarise, this model contains eight physical parameters for each planet, as defined in Section~\ref{sec:transitphysicalmodel} and Section~\ref{sec:rvphysicalmodel}, i.e.\
$P_i$, $T_{0,i}$, $R_{p_i}/R_\star$, $a_i/R_\star$, $b_i$, $e_i$, $\omega_i$, and $M_i$. In addition, we provide $M_\star$ and $R_\star$ to the model (see Section~\ref{sec:transitmodel}), because these values inform $a_i/R_\star$, and because the values are used to calculate derived parameters. A mean flux parameter, $\mu_\mathrm{norm}$ is included, as well as a GP model for the transit light curve as a function of time, as well as a GP model for the RV data. The resulting parameters are $S_0$, $\alpha_0$, $S_1$, $\alpha_1$, $Q_1$, $\sigma_{\mathrm{phot}}$, $\sigma_{2,\mathrm{rv,ESPRESSO}}$, and $\sigma_{2,\mathrm{rv,HARPS}}$, as defined in Section~\ref{sec:transitnoisemodel} and Section~\ref{sec:rvnoisemodel}. A summary of all parameters and their priors is given in Table~\ref{tab:priors}. We infer the optimal solution and its uncertainty using {\tt PyMC3} as built into {\tt exoplanet}. PyMC3 uses a Hamiltonian Monte Carlo scheme to provide a fast inference \citep{exoplanet:pymc3}. As we found the parameter distribution to be symmetric, we report the mean and standard deviation for all parameters in Table~\ref{tab:parameters}. 

\begin{figure*}
	\includegraphics[width=\columnwidth]{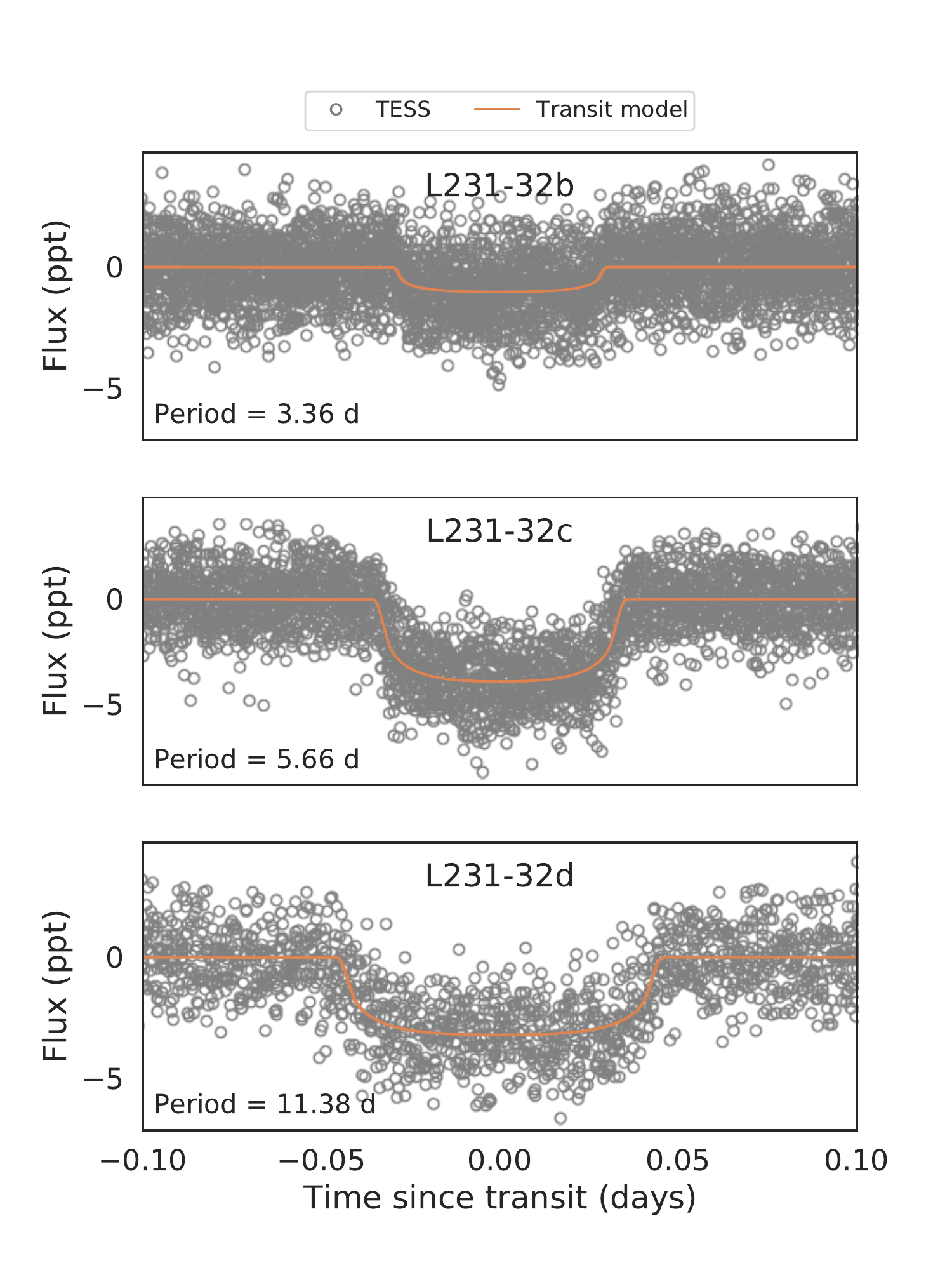}
    \includegraphics[width=\columnwidth]{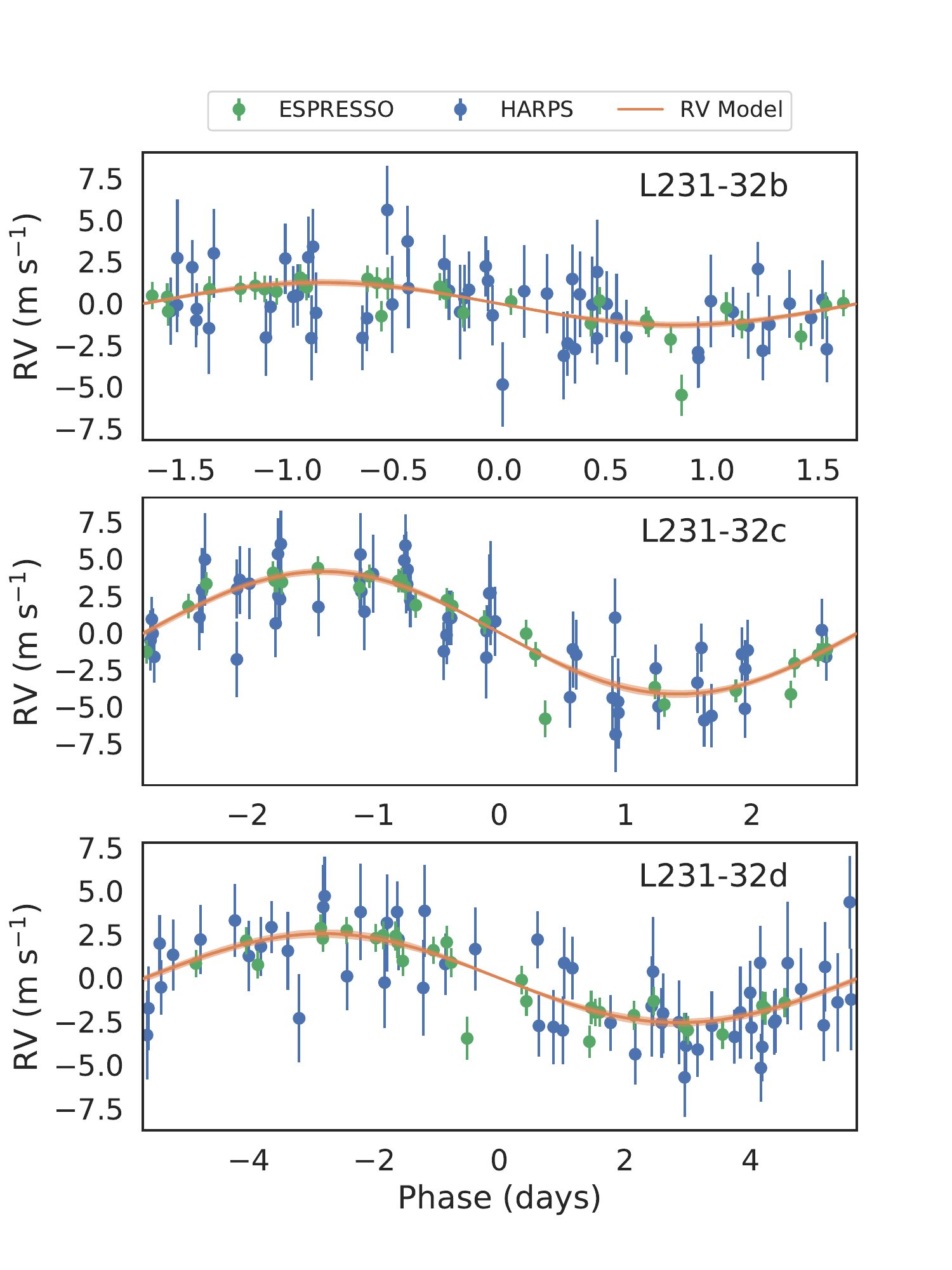}
	\caption{\textit{Left.} The \textit{TESS} light curve folded on the orbital period and centered on the mid-transit time for L231-32b (top), L231-32c (middle) and L231-32d (bottom). The best-fitting model (orange) is shown.
	\textit{Right.} The ESPRESSO (green) and HARPS (blue) data folded on the orbital period for L231-32b (top), L231-32c (middle) and L231-32d (bottom). The best-fitting model (orange) is shown.
	}
	\label{fig:foldeddata}
\end{figure*}

We find that the masses of L231-32b, c, and d, are $1.58 \pm 0.26~M_\oplus$, $6.15~\pm 0.37~M_\oplus$, and $4.78 \pm 0.43~M_\oplus$, respectively.
In Figure~\ref{fig:timeseries} we show the \textit{TESS} photometry together with the best-fitting transit and GP models. We show a zoom-in on the transits folded by orbital period and the RV curve for each planet in Figure~\ref{fig:foldeddata}.

\begin{table*}
  \begin{center}
    \caption{System parameters of L231-32. \label{tab:parameters}}
    \smallskip
    \begin{tabular}{l c c c}
      \hline\hline
      \noalign{\smallskip}
      \multicolumn{3}{c}{Basic properties} \\
      \noalign{\smallskip}
      \hline
      \noalign{\smallskip}
      TESS ID & \multicolumn{2}{c}{TOI-270, TIC 259377017, L231-32} \\
      2MASS ID	 				& \multicolumn{2}{c}{J04333970--5157222} \\
      Right Ascension (hms)				& \multicolumn{2}{c}{04 33 39.72}  	\\ 
      Declination (deg)				& \multicolumn{2}{c}{-51 57 22.44}  	\\
      Magnitude ($V$)				& \multicolumn{2}{c}{$V$: 12.62. 2MASS, $J$: $9.099 \pm 0.032$, $H$: $8.531 \pm 0.073$, $K$: $8.251 \pm 0.029$,}\\
      				& \multicolumn{2}{c}{\textit{TESS}: 10.42, \textit{Gaia}, G: 11.63, b$_p$: 12.87, r$_p$: 10.54}\\
      \hline
      \noalign{\smallskip}
      \multicolumn{3}{c}{Adopted stellar parameters} \\
      \noalign{\smallskip}
      \hline
      \noalign{\smallskip}
      Effective Temperature, $T_{\rm_{eff}}$ (K) 			& \multicolumn{2}{c}{$3506 \pm 70$ }	\\
      Stellar luminosity, $L (\mathrm{L}_\odot$)			& \multicolumn{2}{c}{$0.0194 \pm 0.0019$ }\\ [3pt]
      Surface gravity, $\log g$ (cgs)            			& \multicolumn{2}{c}{$4.872 \pm 0.026$}	\\ [3pt]
      Metallicity, [Fe/H]                        			& \multicolumn{2}{c}{$-0.20 \pm 0.12$ }	\\
      Stellar Mass,   $M_{\star} $ ($M_{\odot}$)			& \multicolumn{2}{c}{$0.386 \pm 0.008$}\\[3pt]
      Stellar Radius, $R_{\star} $ ($R_{\odot}$)	 		& \multicolumn{2}{c}{$0.378 \pm 0.011$} \\[3pt]
      Stellar Density, $\rho_\star$ (g cm$^{-3}$)	 		& \multicolumn{2}{c}{$7.20 \pm 0.63$}	 \\
      Distance (pc)                                         & \multicolumn{2}{c}{$22.453 \pm 0.059$} \\
      
      \hline
       \noalign{\smallskip}
      Parameters from RV and transit fit				& L231-32b		& L231-32c & L231-32d\\
       \noalign{\smallskip}
       \hline
       \noalign{\smallskip}
      Orbital Period, $P$ (days) 					           &  $3.3601538 \pm0.0000048$ 	    & $5.6605731 \pm 0.0000031$           & $11.379573 \pm 0.000013$ \\
      Time of conjunction, $t_{\rm c}$ (BJD$_\mathrm{TDB}-2458385$) 		   &  $2.09505 \pm 0.00074$ 	    & $4.50285 \pm 0.00029$          & $4.68186 \pm 0.00059$ \\ 
      Planetary Mass,   $M_{\rm p} $ ($M_{\oplus}$)			   & $1.58 \pm 0.26$    	& $6.15 \pm 0.37$     & $4.78 \pm 0.43$ \\ 
      Planetary Radius, $R_{\rm p} $ ($R_{\oplus}$)			   & $1.206 \pm 0.039$          & $2.355 \pm 0.064$       & $2.133 \pm 0.058$ \\
      Planetary Density, $\rho_{\rm p}$ (g\,cm$^{-3}$) 		   & $4.97 \pm 0.94$     	      	& $2.60 \pm 0.26$               & $2.72 \pm 0.33$ \\
      Semi-major axis, $a$ (AU)	 				               & $0.03197 \pm 0.00022$  		& $0.04526 \pm 0.00031$     & $0.07210 \pm 0.00050$ \\
      Equilibrium temperature, Ab~$= 0$, $T_\mathrm{eq}$ (K)			   & $581 \pm 14$            	& $488 \pm 12$                  & $387 \pm 10$ \\
      Equilibrium temperature, Ab~$= 0.3$, $T_\mathrm{eq}$ (K)			   & $532 \pm 13$            	& $447 \pm 11$                  & $354 \pm 8$ \\
      Orbital eccentricity, $e$ 					           &  $0.034 \pm 0.025$         & $0.027 \pm 0.021$        		& $0.032 \pm 0.023$\\ 
      Argument of pericenter, $\omega$ (rad)         		   &  $0 \pm 1.8$			& $0.2 \pm 1.6$               & $-0.1 \pm 1.6$\\
      Stellar RV amplitude, $K_\star$ (m\,s$^{-1}$)        	   &  $1.27 \pm 0.21$           & $4.16 \pm 0.24$               & $2.56 \pm 0.23$ \\            
      Fractional Planetary Radius, $R_{\rm p}/R_\star$ 		   & $0.02920 \pm 0.00069$     	    & $0.05701 \pm 0.00071$         & $0.05163 \pm 0.00069$\\
      Impact parameter, $b$ 						           &  $0.19 \pm 0.12$ 		    & $0.28 \pm 0.11$               & $0.19 \pm 0.11$\\
      Inclination, $i$ 						           &  $89.39 \pm 0.37$ 		    & $89.36 \pm 0.24$               & $89.73 \pm 0.16$\\
      Limb darkening parameter, $q_1$					       & $0.17\pm 0.10$ 		    & $0.17\pm 0.10$                & $0.17\pm 0.10$ \\
      Limb darkening parameter, $q_2$ 					       & $0.71\pm 0.16$ 	    	& $0.71\pm 0.16$                & $0.71\pm 0.16$ \\
      \hline
       \noalign{\smallskip}
      \multicolumn{3}{l}{Noise parameters and hyperparameters from RV and transit fit}				\\
       \noalign{\smallskip}
       \hline
       \noalign{\smallskip}
      Photometric `jitter', $\sigma_{\mathrm{phot}}$ (ppt)                          & \multicolumn{2}{c}{$0.5224 \pm 0.0049$} \\ 
      ESPRESSO RV `jitter', $\sigma_{\mathrm{2,rv}}$ (m\,s$^{-1}$)                      & \multicolumn{2}{c}{$0.68 \pm 0.26$} \\ 
      HARPS RV `jitter', $\sigma_{\mathrm{2,rv}}$ (m\,s$^{-1}$)                      & \multicolumn{2}{c}{$0.16 \pm 0.23$} \\ 
      GP power (phot), $S_0$ (ppt$^2$\,days/$2\pi$)                                                         & \multicolumn{2}{c}{$1.29 \pm 0.27$} \\ 
      GP frequency (phot), $\alpha_0$ ($2\pi$/days)                                                & \multicolumn{2}{c}{$1.10 \pm 0.11$} \\ 
      GP power (RV), $S_1$ (m$^2$\,s$^{-2}$\,days/$2\pi$)                                                         & \multicolumn{2}{c}{$22 \pm 70$} \\ 
      GP frequency (RV), $\alpha_1$ ($2\pi$/days)                                                  & \multicolumn{2}{c}{$0.22 \pm 0.36$} \\ 
      GP quality factor Q (RV)                                                                                & \multicolumn{2}{c}{$3.7 \pm 7.3$} \\ 
      ESPRESSO offset, $\gamma_0$ (m~s$^{-1}$)	   	                                   &  \multicolumn{2}{c}{$26850.80 \pm 0.56$} \\ 
      HARPS offset, $\gamma_0$ (m~s$^{-1}$)	   	                                   &  \multicolumn{2}{c}{$26814.28 \pm 0.36$} \\ 
	\hline
      \noalign{\smallskip}
    \end{tabular}
  \end{center}
\end{table*}

\section{The composition of L231-32b, c, and d}
\label{sec:results_toi270}

\subsection{Bulk densities and compositions}

\begin{figure*}
	\includegraphics[width=\textwidth]{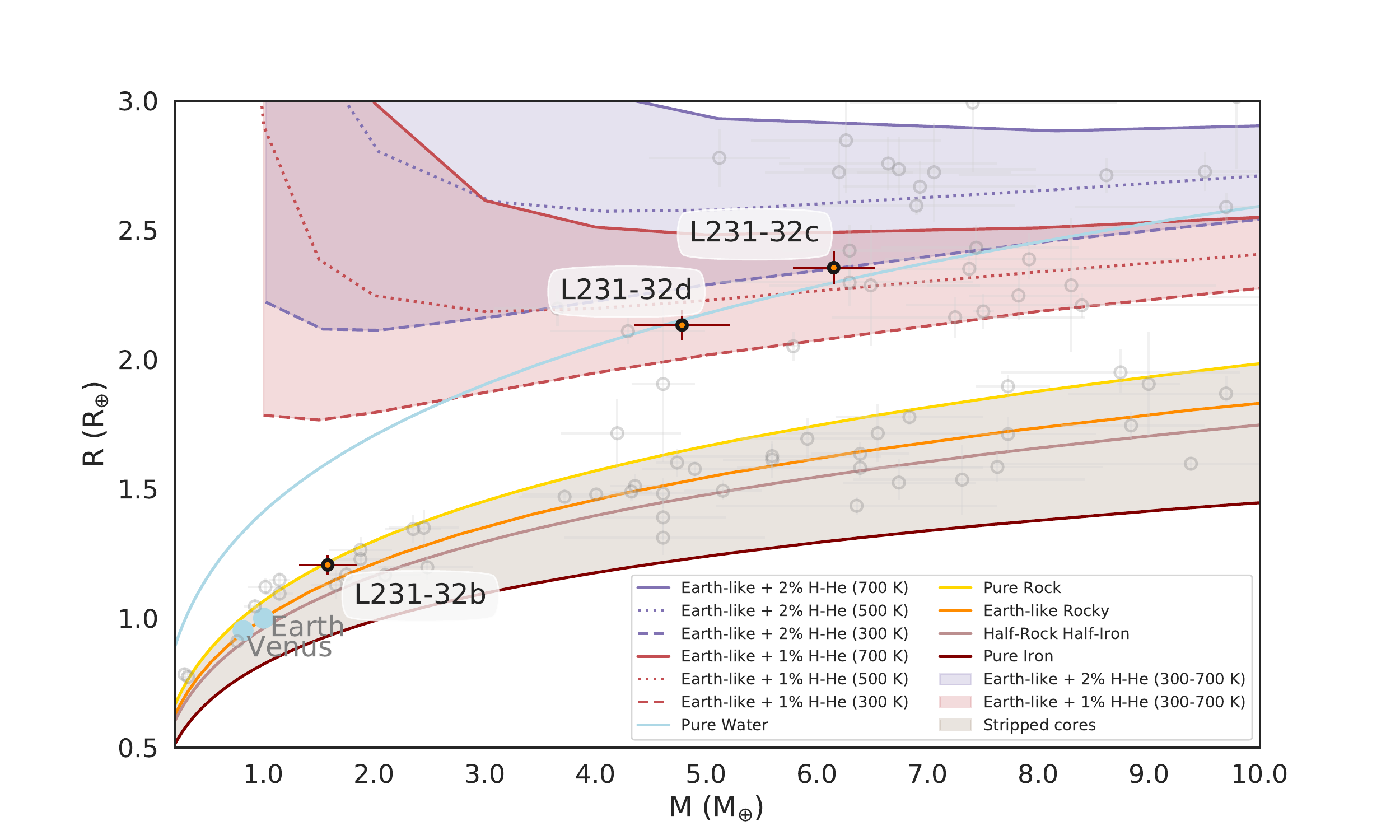}
	\caption{Mass-radius diagram. The planets orbiting L231-32 are indicated in red. Other planets with masses and radii measured to better than 20\% (and $R < 3~R_\oplus$ and $M < 10~M_\oplus$) are shown in grey, with values taken from TEPcat (see Section~\protect\ref{sec:results_toi270} for details). Theoretical lines indicate composition models. Solid lines show models for cores consisting of pure Iron ($100$\% Fe), Earth-like rocky (32.5\%$\mathrm{Fe}$, 67.5\%$\mathrm{MgSiO}_3$), Half-Rock Half-Iron (50\%$\mathrm{Fe}$, 50\%$\mathrm{MgSiO}_3$), and pure Rock (100\%$\mathrm{MgSiO}_3$). A `pure water' model is also shown. In dashed, dotted, and solid lines, models with an Earth-like rocky core and an envelope of H-He taking up 1\% or 2\% of the mass are shown, for temperatures of 300~K, 500~K, and 700~K. All composition models are taken from \protect\cite{zeng2019}. L231-32b is consistent with an Earth-like composition (without a significant envelope). We consider it most likely that L231-32c and L231-32d consist of an Earth-like core composition and a H$_2$ envelope of about 1\% of the planet's total mass, with equilibrium temperatures of around 500 and 300~K, respectively.}
	\label{fig:massradius}
\end{figure*}

\begin{figure*}
	\includegraphics[width=\textwidth]{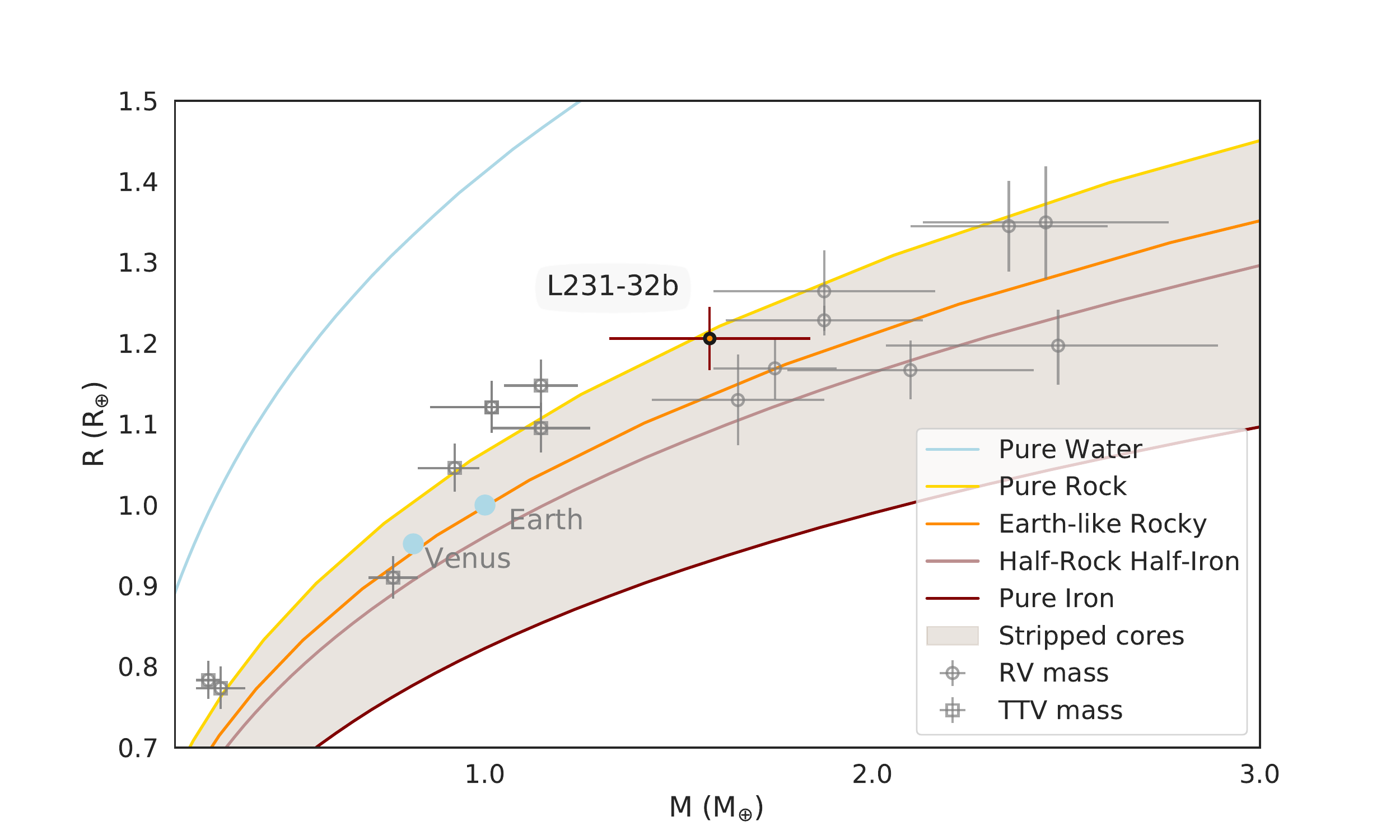}
	\caption{Mass-radius diagram for Earth-mass planets. This figure is similar to Figure~\ref{fig:massradius}, but zoomed in on planets with $M < 3 M_\oplus$. Only a small number of small planets have masses measured to better than 20\%. The seven least massive planets all orbit TRAPPIST-1, and their masses were determined through TTVs. The other planets, in order of increasing mass, are L231-32b, GJ~1132b, LHS 1140c, GJ~3473b, Kepler-78b, GJ~357b, LTT~3780b, L98-59c, and K2-229b. Their masses were determined through RVs. All of these planets follow relatively similar composition tracks, consistent with a composition similar to that of Earth, or slightly more dense (more Iron) or less dense (more Rock). Unlike the other two planets orbiting L231-32, i.e.\ L231-32c and L231-32d, none of these planets have a low density, and they are all inconsistent with a composition of a pure water planet or compositions that would include the presence of a significant H-He atmosphere.
	}
	\label{fig:massradius_zoom}
\end{figure*}

Combining the mass measurements (see Section~\ref{sec:joint}) with the modelled radii ($1.206 \pm 0.039~R_\oplus$, $2.355 \pm 0.064~R_\oplus$, and $2.133 \pm 0.058~R_\oplus$), we find planet densities of $4.97 \pm 0.94$~g~cm$^{-3}$, $2.60 \pm 0.26$~g~cm$^{-3}$, and $2.72 \pm 0.33$~g~cm$^{-3}$, respectively. This implies that the density of the smaller inner planet is significantly higher than that of the two larger, outer planets. 

We now place the mass and radius measurements of the planets orbiting L231-32 into context and compare them to composition models. In Figure~\ref{fig:massradius}, we show a mass-radius diagram for small planets ($R < 3~R_\oplus$ and $M < 10~M_\oplus$). The properties of L231-32 are shown, along with those of other planets for which planet masses and radii are determined to better than 20\%. To do so, we made use of the TEPcat\footnote{\url{https://www.astro.keele.ac.uk/jkt/tepcat/}} database \citep{southworth2011} as a reference, which includes both masses measured through RVs and TTVs. 
We furthermore show composition models taken from \cite{zeng2019}\footnote{Models are available online at \url{https://www.cfa.harvard.edu/~lzeng/planetmodels.html}}. 

As can be seen in Figure~\ref{fig:massradius}, L231-32b is consistent with a composition track corresponding closest to an Earth-like rocky composition (i.e., 32.5\%$\mathrm{Fe}$, 67.5\%$\mathrm{MgSiO}_3$). There are only a few systems with radii as small as that of L231-32b with well-constrained masses. The only lower-mass planets with precisely known masses and radii are the seven planets orbiting TRAPPIST-1 \citep{gillon2016,gillon2017,grimm2018}. Subsequently, L231-32b is now the lowest-mass exoplanet with masses and radii known to better than 20\% with a mass measured through RV observations. In the range of $M_p < 3~M_\oplus$, there are eight other planets with precisely known masses and radii, in order of increasing mass they are 
GJ 1132b \citep{bertathompson2015,bonfils2018}, 
LHS~1140c \citep{dittmann2017,lillobox2020},
GJ 3473b \citep{kemmer2020},
Kepler-78b \citep{pepe2013,howard2013}, 
GJ 357~b \citep[TOI-562;][]{luque2019,jenkins2019}, 
LTT 3780b \citep[TOI-732;][]{nowak2020,cloutier2020ltt3780}, 
L98-59c \citep[TOI-175;][]{kostov2019,cloutier2019}, and 
K2-229b \citep{santerne2018,dai2019}. We zoom in on these small planets in Figure~\ref{fig:massradius_zoom}. From this figure, it is clear that all these planets have a relatively high density, and appear to have a strikingly similar composition, consistent with models with a core composition mixture of MgSiO$_3$ and Fe, similar to Earth, even if some may have a slightly denser (more Iron) or lower density (more rocky) composition. However, all of these planets are inconsistent with lower-density compositions, such as that of pure water planets or planets with even a small mass fraction of H-He atmosphere.

Unlike the TRAPPIST-1 system, where all seven planets have a similar high density, for L231-32 there is a remarkable difference between the density of L231-32b on the one hand, and that of L231-32c and L231-32d on the other. Unlike L231-32b, the two other planets are inconsistent with an Earth-like rocky composition. Instead, when assuming a simple core composition model, the lower density of these planets implies a model such as that of pure water, but it is hard to find a plausible physical reason for why three planets in near-resonant orbits would have formed with such widely different core compositions. We therefore consider an alternative set of models, in which these two outer planets consist not only of a core, but also of a low-density envelope. This atmosphere, which may consist of H-He, can significantly increase the size of a planet even if its contribution to its mass is only minor. 
In Figure~\ref{fig:massradius}, we show composition models \citep[again taken from][]{zeng2019} for an Earth-like core composition (i.e.\ consistent with the composition of L231-32b), as well as a mass fraction of $1\%$ or $2\%$ H-He\footnote{These atmosphere models are referred to as containing H$_2$ by \cite{zeng2019}, but are identical to what is referred to as H-He atmospheres in photo-evaporation models \citep[e.g.][]{owen2013}. Namely, both contain a mixture of H$_2$ and He. Here we use the H-He nomenclature.}. These composition models are sensitive to the effective temperature of the planet. We show models for 300~K, 500~K, and 700~K, as the size of the planet is sensitive to the temperature for a fixed core and atmosphere composition. For L231-32c and L231-32d, we estimate equilibrium temperatures ($T_\mathrm{eq}$) of $447 \pm 11$~K and $354 \pm 8$~K, respectively, assuming an albedo (Ab) of 0.3. As the equilibrium temperature is sensitive to the (unknown) albedo, the true uncertainty is significantly larger, e.g.\ for Ab$~= 0$, we have $T_\mathrm{eq} = 488 \pm 12$ and $T_\mathrm{eq} = 387 \pm 10$ for L231-32c and L231-32d, respectively (see Table~\ref{tab:parameters}).
We find that L231-32c and L231-32d are broadly consistent with models in which their core composition is the same as that of L231-32b (i.e., Earth-like rocky), with the addition of an atmosphere taking up about 1\% of the total mass of the planets, where the precise mass of the H-He envelope is sensitive to assumptions about the exact core composition and equilibrium temperature of these planets. In Section~\ref{sec:mdwarfvalley_theory} we investigate the physical mechanisms that can explain the respective locations of L231-32b, c, and d on the mass-radius diagram in terms of the presence of an H-He atmosphere for the outer planets, and the absence of such an atmosphere for the inner planet.

\subsection{Atmospheric studies of L231-32's planets}

\begin{figure}
	\includegraphics[width=\columnwidth]{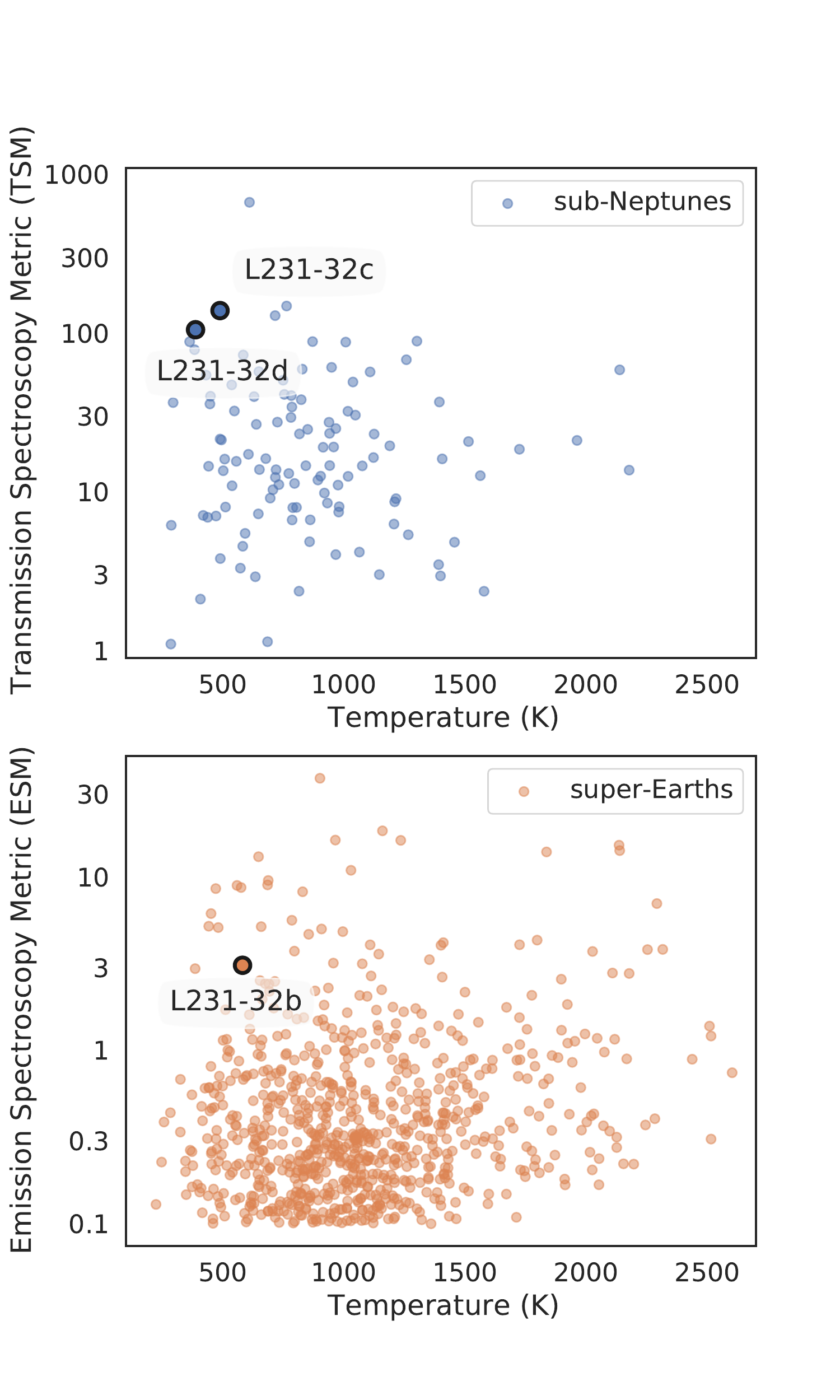}
	\caption{
	Comparison of L231-32 atmospheric metrics against other known exoplanets with $K>5$\,mag. The top panel shows the transmission spectroscopy metric (TSM) values for sub-Neptunes with radii 1.8-4~R$_\oplus$ and published masses, versus planetary equilibrium temperature. L231-32c and L231-32d rank close to the top of all known sub-Neptunes and are currently the most favorable known targets with equilibrium temperatures below 600~K. The bottom panel shows the emission spectroscopy metric (ESM) for super-Earths with radii $R < 1.8~$R$_\oplus$. This includes validated super-Earths without published masses, as the mass does not affect emission spectroscopy. Of the super-Earths, L231-32b ranks among the most favorable with equilibrium temperatures below 1000K.
	}
	\label{fig:atmospheremetrics}
\end{figure}

L231-32c and L231-32d are exciting targets for atmospheric studies for several reasons. First, as we have shown here, L231-32c and L231-32d likely have a significant atmosphere, and determining their atomic and molecular composition will help interpret the evolution history of these planets. L231-32 is an M3V star, with a radius of 0.38~$R_\odot$, which results in relatively deep transits even for small planets, making them more feasible for atmospheric studies through transmission spectroscopy. Additionally, the star is nearby (22 pc) and relatively bright ($K = 8.25$, $V = 12.6$). Indeed, the first atmospheric studies of L231-32c and L231-32d are already ongoing with HST \citep{mikalevans2019}.

Figure~\ref{fig:atmospheremetrics} compares the atmospheric characterisation prospects for the three L231-32 planets with the rest of the known sub-Neptune and super-Earth populations. Specifically, the transmission spectroscopy metric (TSM) and emission spectroscopy metric (ESM) of \cite{kempton2018} have been used to quantify relative signal-to-noises that will be achievable at these two viewing geometries. Planet properties were obtained from the NASA Exoplanet Archive. A brightness cut $K>5$\,mag was applied, as it will be challenging to observe targets brighter than this with JWST \citep[e.g.][]{beichman2014} or using multi-object spectroscopy with large ground-based telescopes such as VLT \citep[e.g.][]{nikolov2018}. 

We consider transmission spectroscopy for the sub-Neptunes, as their low densities make them suitable targets for this type of observation. Of the sub-Neptunes with radii $1.8-4~R_\oplus$ and published masses, L231-32c and L231-32d rank among the most favorable targets (top panel of Figure~\ref{fig:atmospheremetrics}). Indeed, simulations for L231-32c and L231-32d have already shown them to be prime targets for atmospheric studies using JWST \citep{chouqar2020}. Our new mass determinations rule out a water-dominated atmosphere scenario, significantly decreasing the expected number of transit observations necessary for molecular detections, and \cite{chouqar2020} estimate that fewer than three transits with NIRISS and NIRSpec may be enough to reveal molecular features for clear H-He-rich atmospheres.
 
Meanwhile, the super-Earths with radii $<1.8~R_\oplus$ are unlikely to have retained thick H-He-dominated atmospheres. Instead, if they possess significant atmospheres, they are likely to have been outgassed from the interior and to have significantly higher mean molecular weights, making transmission spectroscopy more challenging. \cite{Koll2019} have flagged thermal emission measurements with JWST as a promising alternative method for inferring the presence of an atmosphere on such planets. The bottom panel of Figure~\ref{fig:atmospheremetrics} shows that L231-32b ranks moderately high as a target for this type of measurement, as quantified by its ESM value relative to other super-Earths. It is also worth noting that L231-32b has a relatively low equilibrium temperature among the super-Earths with comparable or higher ESM values. This raises the likelihood that if L231-32b possesses an outgassed atmosphere with a high mean molecular weight, it may have avoided photoevaporative loss, increasing its appeal as a potential rocky target for JWST follow-up observations.

\subsection{Transit timing variations}

As outlined in \cite{guenther2019}, the two outer detected planets, L231-32c and L231-32d, are expected to produce measurable transit timing variations (TTVs) due to their proximity to 5 to 3 (planet $c$ to $b$) and 2 to 1 (planet $d$ to $c$) resonant configurations. The expected TTV period is approximately 1100 days \citep{guenther2019}. Further transit observations using other ground-based or space-based instruments may help constrain the TTV signal of these planets (Kaye et al., in prep.). Such TTV measurements may further refine the planet masses, as well as constrain the orbital eccentricities and arguments of pericenter. 

\section{The radius valley for M dwarf stars}
\label{sec:radiusvalley}

\subsection{The three planets orbiting L231-32 and the radius valley}

As seen in Figure~\ref{fig:massradius}, L231-32b is consistent with a rocky composition without any significant atmosphere. The density of L231-32c and L231-32d is significantly lower. This may suggest a much lower-density core, such as a pure water planet, or a core composition similar to that of L231-32b with a H-He atmosphere. Here, we argue that the latter scenario naturally explains the masses and radii of the three planets orbiting L231-32, and that L231-32b likely formed with an initial H-He envelope similar to that of the two other planets, but that this atmosphere has been lost so that only a stripped core remains.

Although the existence of water worlds has been advocated \citep[e.g.][]{zeng2019}, it is unlikely that the three planets close to mean-motion resonances have different compositions. Specifically, population synthesis models tend to favour the formation of resonant systems that are either all water-poor or water-rich \citep{Izidoro2019,Bitsch2019}; only in rare cases where initial formation straddled the water snow-line could systems with inner rocky planets and outer water-rich planets be formed \citep{Raymond2018}. Alternatively, a model in which L231-32c and L231-32d have a similar, Earth-like rocky, core composition as L231-32b can match its locations in the mass-radius diagram, if one is willing to assume they contain a H-He atmosphere. These atmospheres do not need to be very massive, with a H-He atmosphere of about 1\% of the total planet mass sufficient to explain its mass and radius, as even a tiny mass fraction of a H-He atmosphere significantly increases the planet size (see Figure~\ref{fig:massradius}). The exact planet size for a given H-He envelope mass fraction depends on the temperature of the planet, a quantity which is generally unknown, as it depends on a planet's albedo, which is typically unknown.

A bimodality in the size and composition of small planets has been predicted as a consequence of photo-evaporation in which some planets can lose their entire atmosphere, while others hold on to a H-He envelope \citep[e.g.][]{owen2013,lopez2013}. Planets with a H-He atmosphere, often called sub-Neptunes, are significantly larger in size, than stripped core planets that have lost their atmosphere, i.e.\ the super-Earths. A valley in the radius distribution separating these two types of planets has been observed at about $1.6~R_\oplus$ \citep[e.g.][]{fulton2017,vaneylen2018,fulton2018,berger2018}. The valley's exact location is a function of orbital period \citep{vaneylen2018} and may be largely devoid of planets \citep{vaneylen2018,petigura2020}. Alternative interpretations of the radius valley have been put forward, such as a `core-powered mass loss' scenario in which atmosphere loss of planets is driven by the luminosity of the cooling planet core \citep[e.g.][]{ginzburg2018,gupta2020}. In this scenario, the physical mechanism for atmosphere loss is different, but as in the photo-evaporation scenario the result is a population of stripped core, super-Earth planets that have lost their atmospheres, which is separated by a radius valley from sub-Neptunes, which held on to a H-He envelope.

\begin{figure*}
	\includegraphics[width=\textwidth]{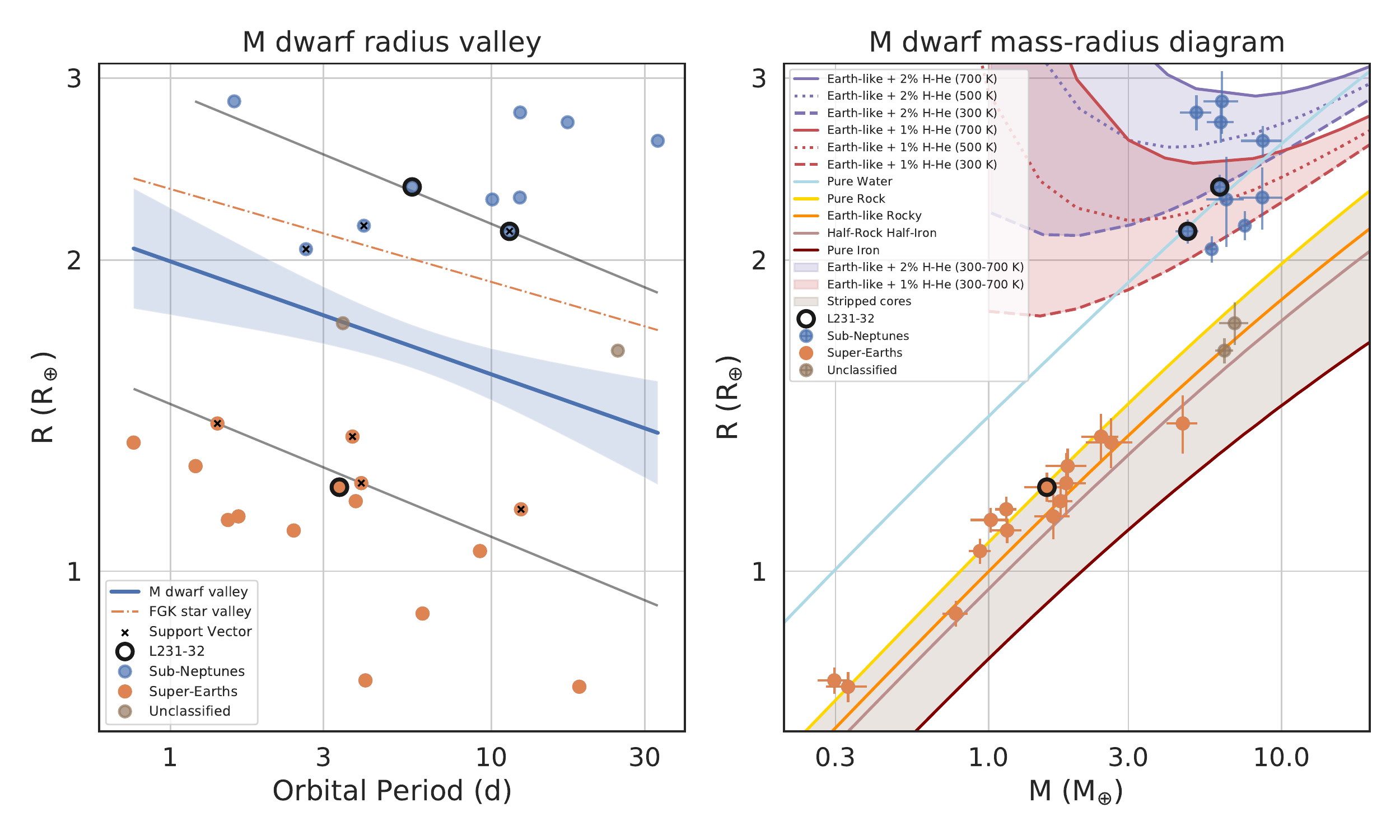}
	\caption{Mass-radius and period-radius diagrams, showing planets orbiting M dwarf stars ($T_\mathrm{eff} < 4000~$K) with masses measured to better than 20\%. On the left, an M dwarf radius valley is shown in blue, determined as the hyperplane of maximum separation using support vector machines, with the grey lines determined by the support vectors. The uncertainty on the location of the valley is determined through bootstrap resampling of the sample of planets. This valley classifies the planets into two categories: super-Earths below the valley and sub-Neptunes above. Two systems, TOI-1235b and LHS 1140b, were not classified. The dash-dotted line shows the location of the radius valley for FGK stars as determined by \protect\cite{vaneylen2018}. On the right, we show the position of the sub-Neptunes and super-Earths on a mass radius diagram together with various composition models similar to Figure~\ref{fig:massradius}. Sub-Neptunes are consistent with a rocky core composition with a $1-2$\% mass H-He atmosphere, whereas super-Earths are consistent with being rocky cores without an atmosphere. This matches thermally driven atmospheric mass loss models.}
	\label{fig:radiusvalley}
\end{figure*}

The location and slope of the radius valley as observed by \cite{vaneylen2018} is consistent with models suggesting planets below the radius valley are stripped cores of terrestrial composition, which have lost their entire atmospheres. Furthermore, the emptiness of the valley would suggest a homogeneity in core composition \citep[e.g.][]{owen2017}. This appears to be what is observed. L231-32b, the size of which suggests it is a super-Earth, located below the valley, is consistent with a terrestrial composition, as are other low-mass planets in the mass-radius diagram (see Figure~\ref{fig:massradius_zoom}). On the other hand, planets with a size on the other side of the valley are predicted to contain a significant H-He atmosphere, which roughly doubles their size but contributes only a small amount of mass \citep[e.g.][]{owen2017}. This is consistent with the observation of L231-32c and L231-32d, the size of which indicates they are sub-Neptunes, located on the upper side of the radius valley. 

We can further check whether the masses and radii of L231-32's planets are quantitatively consistent with photo-evaporation models, in terms of which planets could have lost their atmospheres based on the star's history of XUV flux. However, as this XUV history is not well-understood, we can instead use the \textit{relative} composition of the three planets in this system. Under photo-evaporation models, L231-32b is assumed to have lost its entire initial atmosphere, based on which we can calculate a minimum mass required for L231-32c and L231-32d \textit{not} to have lost their atmospheres. Based on the parameters and their uncertainties listed in Table~\ref{tab:parameters}, we calculate the minimum mass of L231-32c and L231-32d using the {\sc EvapMass} code\footnote{\url{https://github.com/jo276/EvapMass}} as outlined by \cite{owen2020}. These minimum masses answer the question: assuming all planets in the system were born with H-He atmospheres and that L231-32b was stripped of its atmosphere, how massive do L231-32c and L231-32d need to be? We find (at the 95\% confidence level) the photo-evaporation model requires L231-32c to be more massive than 1.04~M$_\oplus$ and L231-32d to be more massive than 0.44~M$_\oplus$. These lower limits are not particularly constraining, and the measured masses are significantly larger than these masses. This can be understood because XUV irradiation is a function of orbital period. As a result, a scenario in which the inner planet loses its atmosphere, while the more distant planets hold on to theirs, is often (though not always) consistent with photo-evaporation.

To summarise, we find here that the mass and radius measurements of the three planets orbiting L231-32 are consistent with photo-evaporation models, in which all three planets started out with a Earth-like rocky core and a H-He envelope. This envelope was retained by L231-32c and L231-32d, but stripped away for L231-32b, placing these planets on opposite `banks' of the radius valley. In Section~\ref{sec:mdwarfvalley_theory}, we investigate what photo-evaporation and core-powered mass loss models predict about the location of the M dwarf radius valley, and in Section~\ref{sec:mdwarfvalley_obs}, we test these models by comparing L231-32 with other well-studied exoplanets orbiting M dwarf stars and determining the location of the M dwarf radius valley.

\subsection{Expected location of the M dwarf radius valley}
\label{sec:mdwarfvalley_theory}

We now set out to empirically determine the location of the radius valley for planets orbiting M dwarf stars. To do so, we first quantify where models predict the M dwarf radius valley to be.
Assuming the photo-evaporation model, we can estimate the position of the the upper edge of the super-Earths (i.e.\ the `lower boundary' of the radius gap). Following \citet{owen2017}, \citet{owen2019}, and \citet{Mordasini2020}, this upper-edge of super-Earths is given by the maximum core size for which photo-evaporation can strip away the atmosphere at that orbital period. This maximum core size can be found by equating the mass-loss timescale $t_{\dot{m}}$ to the saturation timescale of the high-energy output of the star $t_{\rm sat}$; adopting energy limited mass-loss we find (Equation~27 of \citealt{owen2017}):
\begin{equation}
    \frac{GM_p^2 X_2 }{8\pi R_c^3} \sim \frac{\eta}{a_p^2}L^{\rm sat}_{\rm HE}t_{\rm sat} 
\end{equation}
with $a_p$ the planet's semi-major axis, $\eta$ the mass-loss efficiency, $X_2$ the envelope mass-fraction that doubles the core's radius (this is, approximately, the envelope which is hardest for photo-evaporation to strip), and $L^{\rm sat}_{\rm HE}$, the high energy luminosity of star in the saturated phase. The stellar mass-dependence of the radius gap position is encapsulated in how the quantity $L^{\rm sat}_{\rm HE}\times t_{\rm sat} $ varies with stellar mass. Adopting the scalings presented in \citet{owen2017} of $X_2\propto P^{0.08} M_*^{-0.15}M_c^{0.17}$, $M_c\propto R_c^4$ and $\eta\propto R_p/M_c$ we find that the position of the bottom of the radius valley scales as\footnote{A similar scaling can be obtained using the \citet{Mordasini2020} model}:
\begin{equation}
    R^{\rm bot}_{\rm valley} \propto P^{-0.16} M_*^{-0.06} \left(L^{\rm sat}_{\rm HE} t_{\rm sat}\right)^{0.12}.
\end{equation}
Observations indicate that $L^{\rm sat}_{\rm HE}/L_{\rm bol}$ is approximately constant for low-mass stars \citep[e.g.][]{Wright2011}, indicating that $L^{\rm sat}_{\rm HE} \propto L_{\rm bol} \propto M_*^{3.2}$ \citep[e.g.][]{Cox2000}. The scaling of $t_{\rm sat}$ with stellar mass is less certain, although it does increase as the stellar mass decreases \citep[e.g.][]{Selsis2007}. \citet{McDonald2019}'s analysis of the empirical stellar X-ray evolution models of \citet{Jackson2012} suggest that $t_{\rm sat}$ roughly scales like $M_*^{-1}$ from G-dwarfs to M-dwarfs. This implies that the radius valley should scale as:
\begin{equation}
    R^{\rm bot}_{\rm valley} \propto P^{-0.16} M_*^{0.19} \label{eqn:MdwarfValley_phot}
\end{equation}
Equation~\ref{eqn:MdwarfValley_phot} suggests the M-dwarf radius gap should lie at slightly lower planetary radii in the radius-period plane when compared to planets orbiting earlier type stars, while the low power indicates that the location of the valley is not expected to strongly vary with stellar mass. 

We can now predict the location of the radius valley, based on the observed radius valley location for FGK stars. For a sample with masses from around 0.8 to 1.4~$M_\odot$, with a median mass of around $1.1~M_\odot$, \cite{vaneylen2018} determined the location of the radius valley as a function of orbital period as $m = -0.09^{+0.02}_{-0.04}$ and $a=0.37^{+0.04}_{-0.02}$, for $\log_{10} R = m \log_{10} P + a$. 
M dwarf stars span a wide range of masses. The M dwarfs for which we have well-characterised planets span a mass range of roughly $0.1-0.6$~$M_\odot$. Even this wide mass range translates into only a modest spread in the expected location of the radius valley, due to the low mass power in Equation~\ref{eqn:MdwarfValley_phot}, i.e.\ the M dwarf radius valley should be located about 65 to 90\% lower than that of FGK stars, or at a range of $a$ from $0.23^{+0.04}_{-0.02}$ to $0.33^{+0.04}_{-0.02}$. 

One can perform a similar analysis, assuming the radius valley is the result of the core-powered mass-loss mechanism. Combining the results on the period dependence from \citet{Gupta2019} and the stellar mass dependence from  \citet{gupta2020} one finds that\footnote{\citet{gupta2020} argue for a slightly steeper than ZAMS $L-M_*$ relation used in the photoevaporation comparison, as core-powered mass-loss is dominated at older, rather than young, ages, unlike photo-evaporation.}:
\begin{equation}
    R^{\rm bot}_{\rm valley} \propto P^{-0.11} M_*^{0.33} \label{eqn:MdwarfValley_cpml}.
\end{equation}
Such a dependence would imply that the M dwarf radius valley is located about 45 to 80\% lower than that of FGK stars, or with $a$ ranging from $0.17^{+0.04}_{-0.02}$ to $0.30^{+0.04}_{-0.02}$.

\subsection{Observed location of the M dwarf radius valley}
\label{sec:mdwarfvalley_obs}

The ideal sample of planets to determine the radius valley is one with homogeneously derived parameters, as was key to unveiling the radius valley for FGK star \citep[e.g.][]{fulton2017,vaneylen2018}. Unfortunately, such a sample is not readily available for M dwarf stars. We therefore opt to use a sample of \textit{well-studied} planets instead. As before, we start from the TEPcat catalogue, and limit our sample to small planets ($R < 3$~ R$_\oplus$) orbiting M dwarf stars ($T_\mathrm{eff} < 4000$~K) with well-characterised radii (better than 20\%) and masses (also better than 20\%). Limiting our sample to planets with precise masses ensures that each of these planets has been the subject of at least one detailed individual study that has determined both planetary and stellar parameters. We then checked the literature for the most precise set of parameters for each of these planets, and list all of their parameters and sources in Table~\ref{tab:mdwarfs}.

In Figure~\ref{fig:radiusvalley}, we show a period-radius diagram of this sample of planets. A distinct paucity of planets is observed at around $R = 1.6--1.8~R_\oplus$. We determine the location and slope of this valley using support vector machines (SVMs), following the same procedure as outlined in \cite{vaneylen2018}. With SVMs we can determine the so-called `hyperplane of maximum separation' between two populations, which in this case will correspond to an equation of the radius valley as a function of orbital period. 

To do so we use SVC (support vector classification) as part of the Python machine learning package \emph{scikit-learn}. As initial classification, we consider planets to be on the lower side of the M dwarf valley if they are below the known location of the radius valley for FGK stars from \cite{vaneylen2018} lowered by a factor 80\% as predicted based on the mean mass of the stars in this sample ($0.3~M_\odot$) and the scaling for photo-evaporation (see Section~\ref{sec:mdwarfvalley_theory}). We then choose a penalty parameter $C$, which represents a tradeoff between maximising the margin of separation and the tolerance for data mis-classification (a high value of $C$ tolerates less mis-classification). As outlined in \cite{vaneylen2018}, we want a $C$ value in which the location of the valley is primarily determined by the planets nearest to it, and for consistency and to facilitate comparison, we choose the same value as in that work, i.e.\ $C=10$. To obtain a realistic uncertainty on the parameters of the hyperplane, we need to assess to which degree the SVM procedure depends on individual planets in the sample. We therefore repeat the SVM calculation for 5,000 bootstrapped samples, drawn from the sample of planets while allowing replacement. We then take the median and standard deviation as best values and uncertainties.

Following this procedure, we find $a = 0.30^{+0.05}_{-0.06}$ and $m = -0.15^{+0.08}_{-0.05}$. In Figure~\ref{fig:radiusvalley}, we also show a mass-radius diagram of this sample, which shows that super-Earths located below the valley appear consistent with atmosphere-free composition models (`stripped cores') and that most sub-Neptunes located above the valley appear consistent with models of Earth-like rocky cores with a H-He envelope containing $1-2$\% of the total mass. Two systems are of particular interest. TOI-1235b \citep{cloutier2020toi1235}, with a period of 3.4 days and a radius of 1.74 R$_\oplus$ is located near the center of the valley. LHS 1140b \citep{dittmann2017}, with a period of 24.7 days and a radius of 1.64~$R_\oplus$, is located above the valley but with a mass of $6.38 \pm 0.45$~M$_\oplus$ \citep{lillobox2020}, its density is most consistent with not having a meaningful atmosphere. To ensure that neither of these planets is driving the measured location and slope of the valley, we exclude these systems from the sample so they cannot be support vectors. When doing so, we find $a = 0.30^{+0.03}_{-0.05}$ and $m = -0.11^{+0.05}_{-0.04}$. These measurements are remarkably consistent, with a slightly less steep slope and we conservatively adopt these values. In Figure~\ref{fig:radiusvalley}, we show the period-radius and mass-radius diagram for our sample as well as the best-fitted radius valley, the support vectors and lines connecting the support vectors, and composition models similar to those in Figure~\ref{fig:massradius}. In Table~\ref{tab:valleyfit} we list the radius valley location determined here for M dwarf stars, and that for FGK stars from \cite{vaneylen2018}.

\begin{table}
\centering
\scriptsize
\footnotesize
\caption{The location of the radius valley for FGK stars and for M dwarf stars, as described by $\log_{10} R = m \log_{10} P + a$.} \label{tab:valleyfit}
\begin{tabular}{llll}
\hline
Stellar type & Slope $m$ & Intercept $a$ & Reference\\
\hline
FGK & $-0.09^{+0.02}_{-0.04}$               & $0.37^{0.04}_{-0.02}$ & \cite{vaneylen2018} \\[1em]
M & $-0.11^{+0.05}_{-0.04}$     & $0.30^{+0.03}_{-0.05}$ & This work \\
\hline
\end{tabular}
\end{table}

From Figure~\ref{fig:radiusvalley}, we can see that the radius valley determined here is capable of separating small planets orbiting M dwarf stars into two categories: super-Earths located below the valley, consistent with a stripped (rocky) core composition, and sub-Neptunes on the other side of the valley, planets which appear to have a similar, rocky, core, but have held on to their H-He atmosphere which contains about $1-2$\% of the planet's mass. 

This separation of super-Earths and sub-Neptunes in terms of both period-radius and mass-radius space is a remarkably good match to predictions inferred from radius valley models (see Section~\ref{sec:mdwarfvalley_theory}). Furthermore, the radius valley appears to be located at slightly lower radii for M stars relative to FGK stars (see Table~\ref{tab:valleyfit}), which matches the mass dependence predicted by both photo-evaporation models (see Equation~\ref{eqn:MdwarfValley_phot}) and core-powered mass loss models (see Equation~\ref{eqn:MdwarfValley_cpml}). 

While the planets on the other side of the radius valley may also be consistent with low-density core compositions such as that of pure water (see again Figure~\ref{fig:radiusvalley}), we consider this scenario less plausible for several reasons. Firstly, although a mix of rocky and lower density cores may similarly result in a radius valley with two distinct populations \citep[e.g.][]{venturini2020}, several planets have densities so low that even \textit{pure} water planets would be too dense, unless some atmosphere was present. Furthermore, the photo-evaporation or core-powered mass loss scenarios (generally, thermally-driven mass loss) appear to much more naturally explain how multiple planets in the same system can end up with a very different mean density. Under these scenarios, all three planets orbiting L231-32 formed with similar cores and atmospheres, and the observed density difference is caused by the inner planet losing its atmosphere. If the outer two planets in this system were instead low-density cores, i.e.\ `water worlds', it's harder to see why their compositions would be so different given that these three planets orbit in near-resonance. 

Other systems that contain planets on both sides of the valley would further strengthen the thermally-driven mass lost argument, if their mean densities were divergent too. For most other multi-planet systems in our sample, the planets are all located on one side of the valley, e.g.\ K2-146 and Kepler-26 each have two sub-Neptunes, while TRAPPIST-1 has seven super-Earths. LTT~3780 contains one super-Earth and one sub-Neptune, and although the orbital periods are not near resonances, their densities are similarly divergent as expected from a photo-evaporation or core-powered mass loss scenario \citep{cloutier2020ltt3780}. 
LHS 1140 is more puzzling. The system contains two planets at very different periods, i.e.\ 3.8 and 24.7 days (see Table~\ref{tab:mdwarfs}). The inner planet is firmly consistent with being a super-Earth, while the outer planet appears to be located just above the valley for that period, but its mass would suggest it is a super-Earth \citep{lillobox2020}. One possible explanation is that this planet, one of the longest-period planets in our sample, followed a different formation pathway, e.g.\ it may have formed later after the gas disk had already dissipated. At least two other planets (not orbiting M stars), Kepler-100c and Kepler-142c, have been found inconsistent with photo-evaporation models \citep{owen2020}. Finally TOI-1235b appears to be located near or `inside' the radius valley. Given its mass \citep{cloutier2020toi1235}, it is most likely a super-Earth that has lost its atmosphere. As TOI-1235 is one of the most massive stars within our sample, it is possible that the radius valley for this type of star is located at a slightly higher planet radius than for the average star in our sample.

The slope of the radius valley as determined here is different from that determined by \cite{cloutier2020}, who found a radius valley proportional to $F^{-0.060\pm0.025}$ where $F$ is the insolation. As a function of orbital period, this corresponds to a slope with the opposite sign as the one determined here. There are several differences between the approach taken here and that by \cite{cloutier2020}. Firstly, the authors used a significantly larger sample of planets than the one considered here, although one that consists of planets that are generally less well-studied. The approach to finding the valley is also different, as in such a larger but less well-studied sample it is harder to directly separate two separate populations. The authors determine the valley's location by correcting the observed planet sample for completeness to determine a planet occurrence rate, and subsequently determining the location of the peak of `rocky' and `non-rocky' planets from which the location of the valley is determined \citep[see also][for details on this approach]{martinez2019}. As a result, whereas in this work the location of the valley is primarily determined by the planets nearest to it, in \cite{cloutier2020} the valley's location is inferred from the `peak' locations of super-Earth and sub-Neptune planets instead. Finally, \cite{cloutier2020} determine the radius valley as a function of incident flux rather than of orbital period. Here, we opt to use orbital period, because the very small stellar mass-dependence of photo-evaporation models suggest this is the observable with the strongest deterministic power (see Equation~\ref{eqn:MdwarfValley_phot} and Section~\ref{sec:mdwarfvalley_theory}), even when considering a wide range of stellar masses. A larger sample of small, well-studied planets orbiting M dwarf stars, may help resolve this discrepancy, ideally with homogeneously derived stellar and planetary parameters and precise composition measurements.  

\section{Conclusions}
\label{sec:conclusions}

We have measured the masses of three planets orbiting L231-32 using observations from ESPRESSO and HARPS. These planets orbit on both sides of the radius valley, and we find that L231-32b, which is located below the valley, has a significantly higher density than L231-32c and L231-32d, which are located on the other side of the radius valley. We find that L231-32b is a good match to composition models of a planet core stripped of its atmosphere, and consisting of a mixture of rock and iron, similar to Earth. L231-32c and L231-32d have significantly lower densities, and are consistent with a terrestrial-type core combined with a H-He atmosphere taking up only 1-2\% of the mass of these planets. 

These findings are a good match to predictions of thermally driven atmospheric mass loss models, such as photo-evaporation or core-powered mass loss models, in which planets below the radius valley have been stripped of their atmosphere whereas planets above the valley have held on to it. We consider such a scenario more plausible than one in which the core composition of L231-32b is markedly different from those of L231-32c and L231-32d, as these planets orbit near resonances (i.e.\ 5 to 3, and 2 to 1) and likely formed around the same time out of similar material.

Putting L231-32 into context with other planets orbiting M dwarf stars, we determined the presence and location of a radius valley and its slope as a function of orbital period. We found that the valley is located at $\log_{10} R = m \log_{10} P + a$ with $m = -0.11^{+0.05}_{-0.04}$ and $a=0.30^{+0.03}_{-0.05}$. This location is similar to the radius valley around FGK stars. The slope is of the same sign and similar, and the valley is located at slightly smaller planet radii for the same orbital period, as predicted by thermally driven atmospheric mass loss models.
We also investigated the composition of planets orbiting M dwarfs, both below and above the radius valley. We find that planets below the valley are consistent with terrestrial-type cores, without an atmosphere (`super-Earths'), whereas those located above the radius valley are generally consistent with having a H-He envelope of 1-2\% of the total planet mass (`sub-Neptunes').

L231-32b is the smallest planets with a precisely (better than 20\%) measured mass through RV observations, highlighting the potential of \textit{TESS} in finding new transiting planets orbiting the nearest and brightest stars, and of state-of-the-art RV instruments such as ESPRESSO to precisely measure their masses. These precise masses will further facilitate the interpretation of atmospheric measurements of L231-32c and L231-32d, as they are prime targets for transmission spectroscopy using HST and JWST. Finally, as L231-32's planets orbit in near-resonances, the observation of further transits, from ground or in future \textit{TESS} observations, may result in the detection of transit timing variations, which could independently constrain the planet masses and further refine the values reported here.

\footnotetext[31]{International Center for Advanced Studies (ICAS) and ICIFI (CONICET), ECyT-UNSAM, Campus Miguelete, 25 de Mayo y Francia, (1650) Buenos Aires, Argentina}
\footnotetext[32]{MIT Kavli Institute for Astrophysics and Space Research, McNair building, 77 Mass Ave 37-538, Cambridge, MA, 02139}
\footnotetext[33]{European Southern Observatory, Alonso de Cordova 3107, Vitacura, Santiago, Chile}
\footnotetext[34]{NASA Ames Research Center, Moffett Field, CA  94035, USA}
\footnotetext[35]{Astronomical Institute of the Czech Academy of Sciences, Fri\v{c}va 298, 25165, Ond\v{r}ejov, Czech Republic}
\footnotetext[36]{Department of Astronomy, University of Concepción, Chile}
\footnotetext[37]{Space Telescope Science Institute, 3700 San Martin Drive, Baltimore, MD 21218, USA}
\footnotetext[38]{Komaba Institute for Science, The University of Tokyo, 3-8-1 Komaba, Meguro, Tokyo 153-8902, Japan}
\footnotetext[39]{JST, PRESTO, 3-8-1 Komaba, Meguro, Tokyo 153-8902, Japan}
\footnotetext[40]{Astrobiology Center, 2-21-1 Osawa, Mitaka, Tokyo 181-8588, Japan}
\footnotetext[41]{Astronomy Department and Van Vleck Observatory, Wesleyan University, Middletown, CT 06459, USA}
\footnotetext[42]{Department of Earth, Atmospheric and Planetary Sciences, Massachusetts Institute of Technology, Cambridge, MA 02139, USA}
\footnotetext[43]{Department of Aeronautics and Astronautics, MIT, 77 Massachusetts Avenue, Cambridge, MA 02139, USA}
\footnotetext[44]{Centro de Astrobiolog\'ia (CSIC-INTA), Carretera de Ajalvir km 4, E-28850 Torrej\'on de Ardoz, Madrid, Spain}

\section*{Data availability}

This paper includes raw data collected by the \textit{TESS} mission, which are publicly available from the Mikulski Archive for Space Telescopes (MAST, \url{https://archive.stsci.edu/tess}). Observations made with ESPRESSO on the Very Large Telescope and with HARPS at the ESO 3.6m telescope (programs 0102.C-0456, 1102.C-0744, 1102.C-0958, and 1102.C-0339) are publicly available at the ESO archive (\url{http://archive.eso.org/}). All processed data underlying this article are available in the article and in its online supplementary material. The code underlying {\sc EvapMass} analysis for planets c \& d's minimum predicted mass from the photoevaporation model are available at \url{https://github.com/jo276/EvapMassTOI270}.

\section*{Acknowledgements}

We are grateful to Megan Bedell for discussions about the extraction and precision of ESPRESSO radial velocity observations and Laura Kreidberg for discussions about HST observations of this system. Part of this work is done under the framework of the KESPRINT collaboration (\url{http://kesprint.science}). KESPRINT is an international consortium devoted to the characterisation and research of exoplanets discovered with space-based missions. Based on observations made with ESPRESSO on the Very Large Telescope, ESO observing programs 0102.C-0456, 1102.C-0744 and 1102.C-0958 and with HARPS at the ESO 3.6m telescope (program 1102.C-0339). Based on data from the \textit{TESS} satellite. Funding for the \textit{TESS} mission is provided by NASA’s Science Mission directorate. We acknowledge the use of public \textit{TESS} Alert data from pipelines at the \textit{TESS} Science Office and at the \textit{TESS} Science Processing Operations Center. This research has made use of the Exoplanet Follow-up Observation Program website and the NASA Exoplanet Archive, which are operated by the California Institute of Technology, under contract with the National Aeronautics and Space Administration under the Exoplanet Exploration Program. This paper includes data collected by the \textit{TESS} mission, which are publicly available from the Mikulski Archive for Space Telescopes (MAST). 
Resources supporting this work were provided by the NASA High-End Computing (HEC) Program through the NASA Advanced Supercomputing (NAS) Division at Ames Research Center for the production of the SPOC data products.
This work has made use of data from the European Space Agency (ESA) mission {\it Gaia} (\url{https://www.cosmos.esa.int/gaia}), processed by the {\it Gaia} Data Processing and Analysis Consortium (DPAC, \url{https://www.cosmos.esa.int/web/gaia/dpac/consortium}). Funding for the DPAC has been provided by national institutions, in particular the institutions participating in the {\it Gaia} Multilateral Agreement.
This research made use of \textsf{exoplanet} \citep{exoplanet:exoplanet} and its dependencies \citep{exoplanet:astropy13, exoplanet:astropy18,exoplanet:exoplanet, exoplanet:kipping13, exoplanet:luger18, exoplanet:pymc3,exoplanet:theano}.
N. A.-D. acknowledges the support of FONDECYT project 3180063.
B.C.E received support from J.E.O's Royal Society 2020 Enhancement Award. 

J.E.O. is supported by a Royal Society University Research Fellowship. This project has received funding from the European Research Council (ERC) under the European Union’s Horizon 2020 research and innovation programme (PEVAP, Grant agreement No. 853022).
S.A and A.B.J. acknowledge support from the Danish Council for Independent Research, through a DFF Sapere Aude Starting Grant no. 4181-00487B.
M.F, I.G., and C.M.P. gratefully acknowledge the support of the Swedish National Space Agency (DNR 65/19 and 174/18).
This work is supported by JSPS KAKENHI Grant Number 19K14783.
This work was supported by FCT - Funda\c{c}\~ao para a Ci\^encia e a Tecnologia through national funds and by FEDER through COMPETE2020 - Programa Operacional Competitividade e Internacionaliza\c{c}\~ao by these grants: UID/FIS/04434/2019; UIDB/04434/2020; UIDP/04434/2020; PTDC/FIS-AST/32113/2017 \& POCI-01-0145-FEDER-032113; PTDC/FIS-AST/28953/2017 \& POCI-01-0145-FEDER-028953. 
P.K. and J.S. acknowledge the grant INTER-TRANSFER number LTT20015.
SM acknowledges support from the Spanish Ministry of Science and Innovation with the Ramon y Cajal fellowship number RYC-2015-17697 and from the grant number PID2019-107187GB-I00.
J.R.M. acknowledges continuous grants from CNPq, CAPES and FAPERN brazilian agencies.
This work is partly supported by JSPS KAKENHI Grant Numbers JP18H01265 and JP18H05439, and JST PRESTO Grant Number JPMJPR1775.
This work is partly financed by the Spanish Ministry of Economics and Competitiveness through project PGC2018-098153-B-C31.
X.B., X.D., and T.F. acknowledge support from the French National Research Agency in the framework of the Investissements d’Avenir program (ANR- 15-IDEX-02), through the funding of the "Origin of Life" project of the Univ. Grenoble-Alpes.
M.E. acknowledges the support of the DFG priority program SPP 1992 "Exploring  the  Diversity  of  Extrasolar  Planets" (HA  3279/12-1).
This work is made possible by a grant from the John Templeton Foundation. The opinions expressed in this publication are those of the authors and do not necessarily reflect the views of the John Templeton Foundation.
A.S.M. acknowledges financial support from the  Spanish Ministry of Science, Innovation and Universities (MICIU) under the 2019 Juan de la Cierva and AYA2017-86389-P programmes. 
M.D. acknowledges financial support from Progetto Premiale 2015 FRONTIERA (OB.FU. 1.05.06.11) funding scheme of the Italian Ministry of Education, University, and Research.
S.C.C.B. acknowledges support from  Funda\c{c}\~ao para a Ci\^encia e a Tecnologia (FCT) through Investigador FCT contract IF/01312/2014/CP1215/CT0004 and national funds (PTDC/FIS-AST/28953/2017) and by FEDER - Fundo Europeu de Desenvolvimento Regional through COMPETE2020 - Programa Operacional Competitividade e Internacionaliza\c{c}\~ao (POCI-01-0145-FEDER-028953) and through national funds (PIDDAC) by the grant UID/FIS/04434/2019. 
K.W.F.L. and Sz.Cs. acknowledge support by DFG grant RA714/14-1 within the DFG Schwerpunkt SPP 1992, Exploring the Diversity of Extrasolar Planets.



\balance
\bibliographystyle{mnras}
\bibliography{references_asteroseismic} 


\appendix
%
%

%
\begin{table*}
\section{Extra material}
\scriptsize
\caption{{RV observations from ESPRESSO, along with activity indicator measurements, i.e. H$\alpha$, Na D, S-index, and CCF full-width at half-maximum (FWHM). Times are indicated in BJD$_\mathrm{TDB}$.
}
\label{tab:rvdata}
}
\begin{center}
 \begin{tabular}{cccccccccc} 
\hline\hline
\noalign{\smallskip}
Time & RV & $\sigma_{\mathrm{RV}}$ & H$\alpha$ & $\sigma$(H$\alpha$) & Na D & $\sigma$(Na D) & S-index  & $\sigma$(S-index) & FWHM\\
~[BJD] & [m s$^{-1}$] & [m s$^{-1}$] & & & & & & & [m s$^{-1}$] \\
\hline
2458524.575069 & 26854.28 & 0.45 & 0.015181 & 0.000043 & 0.005502 & 0.000044 & 0.512400 & 0.020398 & 5318.30\\
2458524.605313 & 26853.82 & 0.54 & 0.015110 & 0.000052 & 0.005227 & 0.000052 & 0.073070 & 0.041309 & 5319.03\\
2458524.687935 & 26852.24 & 0.50 & 0.015115 & 0.000047 & 0.005493 & 0.000049 & 0.127397 & 0.036766 & 5314.70\\
2458525.713911 & 26843.01 & 1.03 & 0.015582 & 0.000100 & 0.006253 & 0.000111 & -- &         --     & 5304.76\\
2458526.582820 & 26844.54 & 0.41 & 0.014839 & 0.000039 & 0.005813 & 0.000041 & 0.243681 & 0.017530 & 5320.44\\
2458526.658014 & 26843.53 & 0.48 & 0.014809 & 0.000045 & 0.005574 & 0.000047 & 0.099060 & 0.029179 & 5322.66\\
2458527.661693 & 26844.52 & 0.61 & 0.015681 & 0.000061 & 0.005606 & 0.000062 & -- &         --     & 5328.63\\
2458527.691314 & 26846.51 & 0.68 & 0.015261 & 0.000066 & 0.006366 & 0.000074 & 6.383091 & -0.42307 & 5311.29\\
2458528.687739 & 26849.53 & 0.42 & 0.015422 & 0.000041 & 0.005849 & 0.000043 & 0.368796 & 0.022608 & 5320.07\\
2458533.572726 & 26853.13 & 0.38 & 0.014337 & 0.000035 & 0.006911 & 0.000041 & 0.452557 & 0.010843 & 5324.93\\
2458535.638955 & 26856.81 & 0.39 & 0.014581 & 0.000037 & 0.006004 & 0.000040 & 0.267663 & 0.015235 & 5314.61\\
2458536.555663 & 26854.22 & 0.37 & 0.014579 & 0.000035 & 0.006444 & 0.000039 & 0.448926 & 0.011249 & 5319.96\\
2458540.551950 & 26854.43 & 0.38 & 0.013917 & 0.000035 & 0.006201 & 0.000039 & 0.366727 & 0.013546 & 5328.47\\
2458540.608485 & 26854.24 & 0.41 & 0.014069 & 0.000038 & 0.006123 & 0.000043 & 0.326258 & 0.016715 & 5326.46\\
2458546.554326 & 26858.85 & 0.38 & 0.013302 & 0.000034 & 0.006458 & 0.000040 & 0.397390 & 0.011749 & 5336.32\\
2458550.519400 & 26849.29 & 0.40 & 0.013146 & 0.000035 & 0.006651 & 0.000043 & 0.453119 & 0.016023 & 5339.77\\
2458550.589370 & 26849.32 & 0.49 & 0.013308 & 0.000044 & 0.006793 & 0.000054 & 0.322875 & 0.034426 & 5335.55\\
2458552.543476 & 26850.77 & 0.43 & 0.013760 & 0.000040 & 0.006773 & 0.000047 & 0.438748 & 0.016620 & 5341.98\\
2458553.534623 & 26850.79 & 0.44 & 0.013012 & 0.000039 & 0.006480 & 0.000048 & 0.379358 & 0.015756 & 5332.79\\
2458555.528411 & 26847.70 & 0.35 & 0.013262 & 0.000032 & 0.007052 & 0.000039 & 0.520167 & 0.010136 & 5334.72\\
2458556.517205 & 26851.48 & 0.40 & 0.013484 & 0.000036 & 0.006877 & 0.000043 & 0.429062 & 0.015259 & 5337.23\\
2458557.520381 & 26859.44 & 0.38 & 0.013955 & 0.000035 & 0.007297 & 0.000042 & 0.630267 & 0.014137 & 5333.91\\
2458557.551631 & 26859.05 & 0.36 & 0.013652 & 0.000032 & 0.007158 & 0.000039 & 0.492472 & 0.012044 & 5341.48\\
2458558.513560 & 26857.33 & 0.40 & 0.014048 & 0.000037 & 0.007521 & 0.000046 & 0.682517 & 0.016662 & 5338.92\\
2458559.526363 & 26851.23 & 0.63 & 0.014045 & 0.000059 & 0.007471 & 0.000073 & 0.534787 & 0.084700 & 5330.13\\
2458559.599837 & 26850.31 & 0.46 & 0.013606 & 0.000042 & 0.007177 & 0.000052 & 0.274579 & 0.030308 & 5343.26\\
2458562.509611 & 26850.06 & 0.45 & 0.013606 & 0.000041 & 0.007303 & 0.000050 & 0.501180 & 0.016013 & 5331.84\\
2458564.558099 & 26853.24 & 0.43 & 0.013919 & 0.000039 & 0.006860 & 0.000046 & 0.377530 & 0.016215 & 5339.88\\
2458564.601209 & 26852.65 & 0.64 & 0.013706 & 0.000058 & 0.007262 & 0.000073 & 0.166569 & 0.033266 & 5328.74\\
\end{tabular}
\end{center}
\end{table*}

%
\begin{table*}
\scriptsize
\caption{{Radial velocity observations from HARPS, similar to Table~\ref{tab:rvdata}. Times are indicated in BJD$_\mathrm{TDB}$, converted for consistency with ESPRESSO and TESS observations.}
\label{tab:rvdata_harps}
}
\begin{center}
 \begin{tabular}{ cccccccccccccc } 
\hline\hline
\noalign{\smallskip}
Time & RV & $\sigma_{\mathrm{RV}}$ & H$\alpha$ & $\sigma$(H$\alpha$) & Na D & $\sigma$(Na D) & S-index & $\sigma$(S-index) & FWHM\\
~[BJD] & [m s$^{-1}$] & [m s$^{-1}$] & & & & & & & [m s$^{-1}$] \\
\hline
2458501.638740 & 26817.89 & 2.62 & 0.064654 & 0.000312 & 0.007010 & 0.000308 & 0.606371 & 0.129219 & 3100.92\\
2458502.606327 & 26815.47 & 1.77 & 0.066337 & 0.000220 & 0.007383 & 0.000178 & 0.629866 & 0.074567 & 3112.36\\
2458503.602859 & 26791.27 & 1.73 & 0.064734 & 0.000205 & 0.007617 & 0.000170 & 0.503732 & 0.088541 & 3089.70\\
2458504.631670 & 26812.37 & 1.80 & 0.065824 & 0.000218 & 0.007632 & 0.000180 & 0.570561 & 0.079752 & 3097.04\\
2458505.632936 & 26809.90 & 1.75 & 0.065485 & 0.000217 & 0.007332 & 0.000173 & 0.572624 & 0.075682 & 3102.44\\
2458506.629421 & 26815.43 & 1.59 & 0.065414 & 0.000192 & 0.006944 & 0.000149 & 0.521618 & 0.090782 & 3096.22\\
2458507.623881 & 26819.91 & 2.13 & 0.065475 & 0.000266 & 0.007041 & 0.000231 & 0.660550 & 0.103650 & 3105.15\\
2458515.711419 & 26804.57 & 2.14 & 0.065083 & 0.000229 & 0.006566 & 0.000215 & 0.805628 & 0.235556 & 3066.80\\
2458516.619701 & 26808.93 & 1.60 & 0.067145 & 0.000192 & 0.006836 & 0.000148 & 0.552813 & 0.111421 & 3083.65\\
2458517.707965 & 26813.62 & 2.41 & 0.066315 & 0.000257 & 0.006271 & 0.000258 & 0.421222 & 0.311990 & 3058.20\\
2458518.595738 & 26811.26 & 1.55 & 0.066935 & 0.000197 & 0.006917 & 0.000146 & 0.471438 & 0.082643 & 3104.18\\
2458518.687807 & 26812.36 & 2.64 & 0.067388 & 0.000311 & 0.007334 & 0.000305 & 0.366182 & 0.242884 & 3087.91\\
2458519.656909 & 26810.61 & 2.35 & 0.067585 & 0.000274 & 0.007298 & 0.000261 & 0.436994 & 0.204376 & 3078.22\\
2458520.611601 & 26805.74 & 2.55 & 0.067720 & 0.000297 & 0.006571 & 0.000285 & 0.379174 & 0.197455 & 3082.92\\
2458520.633776 & 26807.23 & 2.42 & 0.067497 & 0.000284 & 0.006697 & 0.000268 & 0.444928 & 0.184293 & 3086.50\\
2458522.595277 & 26813.28 & 1.50 & 0.067618 & 0.000181 & 0.006113 & 0.000129 & 0.374242 & 0.113518 & 3089.23\\
2458524.596870 & 26818.87 & 1.75 & 0.068818 & 0.000219 & 0.006300 & 0.000168 & 0.429766 & 0.135771 & 3085.08\\
2458524.619462 & 26817.19 & 1.78 & 0.068179 & 0.000217 & 0.006221 & 0.000169 & 0.536764 & 0.147592 & 3091.24\\
2458528.654407 & 26812.28 & 2.88 & 0.068592 & 0.000337 & 0.006683 & 0.000339 & 0.407716 & 0.298258 & 3082.28\\
2458528.676756 & 26814.34 & 3.13 & 0.068547 & 0.000365 & 0.006871 & 0.000387 & 0.307021 & 0.316033 & 3078.01\\
2458530.606801 & 26814.20 & 1.84 & 0.069083 & 0.000226 & 0.006309 & 0.000179 & 0.545872 & 0.142241 & 3084.00\\
2458530.628745 & 26814.24 & 1.84 & 0.069017 & 0.000224 & 0.006935 & 0.000181 & 0.503089 & 0.155916 & 3085.79\\
2458538.618194 & 26808.39 & 1.95 & 0.066202 & 0.000231 & 0.006459 & 0.000195 & 0.728370 & 0.199296 & 3081.59\\
2458538.640334 & 26812.28 & 2.06 & 0.066435 & 0.000242 & 0.006734 & 0.000209 & 0.484243 & 0.198056 & 3075.34\\
2458540.558705 & 26814.78 & 2.29 & 0.066362 & 0.000280 & 0.006913 & 0.000250 & 0.303395 & 0.181834 & 3084.11\\
2458540.580649 & 26816.66 & 1.88 & 0.066530 & 0.000229 & 0.006735 & 0.000189 & 0.528892 & 0.173109 & 3086.57\\
2458541.604011 & 26817.91 & 1.83 & 0.065992 & 0.000219 & 0.006594 & 0.000178 & 0.760144 & 0.179542 & 3081.04\\
2458541.625735 & 26815.79 & 1.81 & 0.065906 & 0.000210 & 0.006525 & 0.000172 & 0.679760 & 0.190828 & 3079.24\\
2458543.572115 & 26814.05 & 1.62 & 0.065838 & 0.000201 & 0.006545 & 0.000148 & 0.512478 & 0.129675 & 3090.80\\
2458543.593839 & 26811.56 & 1.56 & 0.065324 & 0.000187 & 0.006775 & 0.000137 & 0.618684 & 0.137910 & 3089.11\\
2458545.615206 & 26817.88 & 2.23 & 0.074901 & 0.000275 & 0.010693 & 0.000243 & 1.087332 & 0.237695 & 3078.05\\
2458546.558271 & 26819.42 & 1.97 & 0.065589 & 0.000241 & 0.006981 & 0.000199 & 0.674405 & 0.159411 & 3096.96\\
2458548.601664 & 26813.88 & 2.38 & 0.065453 & 0.000292 & 0.007284 & 0.000262 & 0.335639 & 0.197755 & 3086.08\\
2458549.594890 & 26812.40 & 1.64 & 0.065366 & 0.000193 & 0.007379 & 0.000149 & 0.592066 & 0.170009 & 3078.90\\
2458549.616626 & 26807.49 & 1.78 & 0.065142 & 0.000208 & 0.007304 & 0.000168 & 0.535322 & 0.188159 & 3082.12\\
2458551.572813 & 26815.78 & 2.00 & 0.065141 & 0.000233 & 0.007152 & 0.000196 & 0.633940 & 0.205005 & 3078.08\\
2458551.594919 & 26816.34 & 2.29 & 0.064669 & 0.000265 & 0.007168 & 0.000236 & 0.576712 & 0.224159 & 3085.51\\
2458553.582392 & 26816.38 & 3.53 & 0.065129 & 0.000424 & 0.007189 & 0.000445 & 0.711391 & 0.296712 & 3096.95\\
2458554.570747 & 26816.66 & 2.67 & 0.065401 & 0.000319 & 0.007463 & 0.000299 & 0.632174 & 0.216452 & 3091.94\\
2458554.594230 & 26810.98 & 2.92 & 0.065041 & 0.000348 & 0.007328 & 0.000340 & 0.743993 & 0.234926 & 3090.76\\
2458555.603071 & 26812.48 & 1.98 & 0.065508 & 0.000227 & 0.007183 & 0.000191 & 0.565103 & 0.190737 & 3080.90\\
2458556.564945 & 26816.44 & 1.69 & 0.065580 & 0.000202 & 0.007293 & 0.000154 & 0.555501 & 0.141030 & 3092.96\\
2458557.558810 & 26824.16 & 2.43 & 0.066040 & 0.000288 & 0.007333 & 0.000258 & 0.701496 & 0.201289 & 3090.65\\
2458557.581460 & 26824.84 & 2.26 & 0.065645 & 0.000262 & 0.007248 & 0.000232 & 0.727598 & 0.193947 & 3090.21\\
2458558.574735 & 26821.12 & 2.78 & 0.066872 & 0.000329 & 0.007223 & 0.000313 & 0.495422 & 0.238121 & 3100.14\\
2458564.534128 & 26813.34 & 1.95 & 0.066154 & 0.000233 & 0.007373 & 0.000189 & 0.711700 & 0.153764 & 3087.06\\
2458564.555655 & 26814.44 & 1.97 & 0.065907 & 0.000232 & 0.007504 & 0.000191 & 0.519325 & 0.153629 & 3086.72\\
2458565.534847 & 26809.37 & 2.06 & 0.066249 & 0.000238 & 0.007443 & 0.000200 & 2.563182 & 0.187104 & 3084.59\\
2458565.557011 & 26812.60 & 2.58 & 0.066306 & 0.000297 & 0.007829 & 0.000277 & 0.582548 & 0.229589 & 3092.53\\
2458566.544711 & 26811.76 & 2.04 & 0.065955 & 0.000237 & 0.007051 & 0.000193 & 0.473000 & 0.191530 & 3089.12\\
2458567.531126 & 26818.34 & 2.11 & 0.066022 & 0.000253 & 0.007158 & 0.000207 & 0.545315 & 0.157097 & 3086.21\\
2458568.554022 & 26815.68 & 2.54 & 0.067969 & 0.000305 & 0.007521 & 0.000270 & 0.740052 & 0.216288 & 3090.83\\
2458569.533969 & 26821.42 & 2.78 & 0.067216 & 0.000333 & 0.007047 & 0.000304 & 0.536221 & 0.219075 & 3091.80\\
2458570.532156 & 26815.44 & 2.75 & 0.066919 & 0.000330 & 0.007220 & 0.000302 & 0.831315 & 0.230372 & 3094.29\\
2458570.554736 & 26819.80 & 2.65 & 0.066328 & 0.000312 & 0.007377 & 0.000282 & 0.842366 & 0.212231 & 3080.04\\
2458586.511092 & 26813.86 & 1.99 & 0.070205 & 0.000238 & 0.006010 & 0.000183 & 0.495104 & 0.167436 & 3075.81\\
2458588.514878 & 26809.27 & 2.83 & 0.069249 & 0.000328 & 0.006511 & 0.000306 & 0.617578 & 0.257318 & 3070.19\\
2458590.512139 & 26815.60 & 2.03 & 0.069332 & 0.000230 & 0.006550 & 0.000189 & 0.625750 & 0.203256 & 3065.81\\
\end{tabular}
\end{center}
\end{table*}

\begin{table*}
\centering
\scriptsize
\caption{Joint transit and RV model priors. Values and the choice of prior is described in more detail in Section~\protect\ref{sec:planet}. A Normal distribution with mean $\mu$ and standard deviation $\sigma$ is indicated as  $\mathcal{N}(\mu,\sigma)$. A Bounded Normal distribution, truncated by a lower limit $l$ and an upper limit $u$ is indicated as $\mathcal{BN}(\mu,\sigma;l,u)$. A uniform distribution with lower limit $l$ and upper limit $u$ is indicated by $\mathcal{U}(l,u)$. A Beta distribution with parameters $a$ and $b$ is indicated with $\mathcal{B}(a,b)$, if it is further bound by a lower limit $l$ and upper limit $u$, it is indicated as $\mathcal{BB}(a,b;l,u)$.
\label{tab:priors}}
\begin{tabular}{lllr}
\hline
Description & Parameter & Unit & Prior \\ 
\hline
\noalign{\smallskip}
  Stellar mass              & $M_\star$     & $(M_\odot)$               & $\mathcal{BN}(0.386,0.008;0,3)$ \\
  Stellar radius            & $R_\star$     & $(R_\odot)$               & $\mathcal{BN}(0.378,0.011;0,3)$ \\
  Limb darkening            & $q_1$         &                           & $\mathcal{U}(0,1)$ \\
  Limb darkening            & $q_2$         &                           & $\mathcal{U}(0,1)$ \\
  Orbital period (b)        & $P_{b}$       & days                      & $\mathcal{U}(3.35,3.37)$  \\ 
  Orbital period (c)        & $P_{c}$       & days                      & $\mathcal{U}(5.65,5.67)$  \\
  Orbital period (d)        & $P_{d}$       & days                      & $\mathcal{U}(11.37,11.39)$  \\
  Time of conjunction (b)   &$t_{\rm c,b}$ & (BJD$_\mathrm{TDB}-2458387$)            & $\mathcal{U}(0.04,0.14)$ \\
  Time of conjunction (c)   &$t_{\rm c,c}$ & (BJD$_\mathrm{TDB}-2458387$)            & $\mathcal{U}(2.45,2.55)$ \\
  Time of conjunction (d)   &$t_{\rm c,d}$ & (BJD$_\mathrm{TDB}-2458387$)            & $\mathcal{U}(2.63,2.73)$ \\
  Eccentricity (b)          & $e_{\rm b}$   &                           & $\mathcal{BB}(1.52,29.0;0,1)$  \\
  Eccentricity (c)          & $e_{\rm c}$   &                           & $\mathcal{BB}(1.52,29.0;0,1)$  \\
  Eccentricity (d)          & $e_{\rm d}$   &                           & $\mathcal{BB}(1.52,29.0;0,1)$  \\
  Argument of pericenter (b)&$\omega_{\rm b}$ & rad                     & $\mathcal{U}(-\pi,\pi)$  \\ 
  Argument of pericenter (c)&  $\omega_{\rm c}$ & rad                   & $\mathcal{U}(-\pi,\pi)$  \\
  Argument of pericenter (d)&  $\omega_{\rm d}$ &rad                    & $\mathcal{U}(-\pi,\pi)$ \\
  Planet mass (b)           & $M_{\rm p,b}$     & $M_{\oplus}$	        & $\mathcal{N}(2,10)$ \\
  Planet mass (c)           & $M_{\rm p,c}$     & $M_{\oplus}$	        & $\mathcal{N}(7,10)$ \\
  Planet mass (d)           & $M_{\rm p,d}$     & $M_{\oplus}$      	& $\mathcal{N}(3.5,10)$ \\
  Planet radius (b)         & $R_{\rm p,b}$     & $R_{\oplus}$          & $\mathcal{N}(0.0123,10)$ \\
  Planet radius (c)         & $R_{\rm p,c}$     & $R_{\oplus}$          & $\mathcal{N}(0.0231,10)$ \\
  Planet radius (d)         & $R_{\rm p,d}$     & $R_{\oplus}$          & $\mathcal{N}(0.0190,10)$ \\
  Impact parameter (b)      & $b_b$             &                       & $\mathcal{U}(0,1)$ \\ 
  Impact parameter (c)      & $b_c$             &                       & $\mathcal{U}(0,1)$ \\
  Impact parameter (d)      & $b_d$             &                       & $\mathcal{U}(0,1)$ \\ 
  Mean flux                 & mf                &                       & $\mathcal{N}(0,10)$ \\
  Log photometric jitter    & $\log \sigma_{\mathrm{phot}}$ &       & $\mathcal{N}(2,100)$ \\ 
  Offset HARPS              & $\gamma_\mathrm{harps}$ & m~s$^{-1}$      & $\mathcal{N}(0,10)$\\
  Offset ESPRESSO           & $\gamma_\mathrm{espresso}$ & m~s$^{-1}$   & $\mathcal{N}(0,10)$\\
  Log jitter HARPS              & $\log \sigma_{\mathrm{2,rv,HARPS}}$ & m\,s$^{-1}$&  $\mathcal{N}(\log(1),5)$ \\ 
  Log jitter ESPRESSO           & $\log \sigma_{\mathrm{2,rv,ESPRESSO}}$ &m\,s$^{-1}$& $\mathcal{N}(\log(1),5)$ \\
  Photometry hyperparameter & $\log S_0$            &                       & $\mathcal{N}(\log(\sigma_{\mathrm{phot}}^2),10)$ \\
  Photometry hyperparameter & $\log \alpha_0$       &                       & $\mathcal{N}(\log(2\pi/10),10)$ \\
  RV hyperparameter         & $\log S_1$                &                   & $\mathcal{N}(\log(5),2)$ \\ 
  RV hyperparameter         & $\log \alpha_1$           &                   & $\mathcal{N}(\log(2\pi/50),2)$ \\ 
  RV hyperparameter         & $\log Q$                 &                   & $\mathcal{N}(\log(5),2)$ \\
\hline
\end{tabular}
\end{table*}

\begin{table*}
\setlength{\tabcolsep}{4pt}
\scriptsize
\caption{Small planets ($R < 3~R_\oplus$) with well-measured masses ($< 20\%$) and radii ($< 20\%$) orbiting M dwarf ($T_\mathrm{eff} < 4000~K$) stars. For each system, the source of the planet discovery and of each parameter is listed. Some uncertainties were symmetrised for simplicity. \label{tab:mdwarfs}}
\textbf{Notes.} 
(a) \cite{luque2019};
(b)	\cite{bertathompson2015};
(c)	\cite{bonfils2018};
(d)	\cite{charbonneau2009};
(e)	\cite{harpsoe2013};
(f) \cite{kemmer2020};
(g) \cite{crossfield2015};\\
(h)	\cite{kosiarek2019};
(i)	\cite{damasso2018};
(j)	\cite{montet2015};
(k)	\cite{benneke2019};
(l)	\cite{hirano2018};
(m) \cite{hamann2019};
(n)	\cite{steffen2012};
(o)	\cite{jontofhutter2016};\\
(p)	\cite{astudillodefru2020};
(q)	\cite{kostov2019};
(r)	\cite{cloutier2019};
(s)	\cite{dittmann2017};
(t) \cite{lillobox2020};
(u)	\cite{cloutier2020ltt3780};\\
(v) \cite{cloutier2020toi1235};
(w)	\cite{gillon2016};
(z)	\cite{gillon2017};
(y)	\cite{grimm2018};
(z)	This work.
\begin{tabular}{llrlrlrlrlrlrl}
\hline
System		&	Disc.	&	Period			&	Ref.	&	Radius			&	Ref.	&	Mass			&	Ref.	&	Density			&	Ref.	&	Mstar			&	Ref.	&	Rstar			&	Ref.	\\
    		&	    	&	[d]			&		&	[R$_\oplus$]			&		&	[M$_\oplus$]			&		&	[g~cm$^{-3}$]			&	&	[M$_\odot$]			& 	&	[R$_\odot$]			&		\\
\hline
\noalign{\smallskip}
GJ 357	b	&	(a)	&	3.93072	$\pm$	0.00008	&	(a)	&	1.217	$\pm$	0.084	&	(a)	&	1.84	$\pm$	0.31	&	(a)	&	5.6	$\pm$	1.5	&	(a)	&	0.342	$\pm$	0.011	&	(a)	&	0.337	$\pm$	0.015	&	(a)	\\
GJ 1132	b	&	(b)	&	1.628931	$\pm$	0.000027	&	(c)	&	1.13	$\pm$	0.056	&	(c)	&	1.66	$\pm$	0.23	&	(c)	&	6.3	$\pm$	1.3	&	(c)	&	0.181	$\pm$	0.019	&	(c)	&	0.2105	$\pm$	0.0094	&	(c)	\\
GJ 1214	b	&	(d)	&	1.58040456	$\pm$	0.00000016	&	(e)	&	2.85	$\pm$	0.2	&	(e)	&	6.26	$\pm$	0.86	&	(e)	&	1.49	$\pm$	0.33	&	(e)	&	0.15	$\pm$	0.011	&	(e)	&	0.216	$\pm$	0.012	&	(e)	\\
GJ 3473	b	&	(f)	&	1.1980035	$\pm$	0.0000018	&	(f)	&	1.264	$\pm$	0.05	&	(f)	&	1.86	$\pm$	0.3	&	(f)	&	5.03	$\pm$	1	&	(f)	&	0.36	$\pm$	0.016	&	(f)	&	0.364	$\pm$	0.016	&	(f)	\\
K2-3	b	&	(g)	&	10.054626	$\pm$	0.000011	&	(h)	&	2.29	$\pm$	0.23	&	(i)	&	6.48	$\pm$	0.96	&	(h)	&	3.7	$\pm$	1.38	&	(h)	&	0.601	$\pm$	0.089	&	(h)	&	0.561	$\pm$	0.068	&	(h)	\\
K2-18	b	&	(j)	&	32.940045	$\pm$	0.00001	&	(k)	&	2.61	$\pm$	0.087	&	(k)	&	8.63	$\pm$	1.35	&	(k)	&	2.67	$\pm$	0.5	&	(k)	&	0.4951	$\pm$	0.0043	&	(k)	&	0.4445	$\pm$	0.0148	&	(k)	\\
K2-146	b	&	(l)	&	2.6446	$\pm$	0.00006	&	(m)	&	2.05	$\pm$	0.06	&	(m)	&	5.77	$\pm$	0.18	&	(m)	&	3.69	$\pm$	0.21	&	(m)	&	0.331	$\pm$	0.009	&	(m)	&	0.33	$\pm$	0.01	&	(m)	\\
K2-146	c	&	(m)	&	4.00498	$\pm$	0.00011	&	(m)	&	2.16	$\pm$	0.07	&	(m)	&	7.49	$\pm$	0.24	&	(m)	&	3.92	$\pm$	0.27	&	(m)	&	0.331	$\pm$	0.009	&	(m)	&	0.33	$\pm$	0.01	&	(m)	\\
Kepler-26	b	&	(n)	&	12.28	$\pm$	0.0003	&	(o)	&	2.78	$\pm$	0.11	&	(o)	&	5.12	$\pm$	0.63	&	(o)	&	1.26	$\pm$	0.2	&	(o)	&	0.544	$\pm$	0.025	&	(o)	&	0.512	$\pm$	0.017	&	(o)	\\
Kepler-26	c	&	(n)	&	17.2559	$\pm$	0.0006	&	(o)	&	2.72	$\pm$	0.12	&	(o)	&	6.2	$\pm$	0.65	&	(o)	&	1.61	$\pm$	0.25	&	(o)	&	0.544	$\pm$	0.025	&	(o)	&	0.512	$\pm$	0.017	&	(o)	\\
L168-9	b	&	(p)	&	1.4015	$\pm$	0.00018	&	(p)	&	1.39	$\pm$	0.09	&	(p)	&	4.6	$\pm$	0.56	&	(p)	&	9.6	$\pm$	2.1	&	(p)	&	0.62	$\pm$	0.03	&	(p)	&	0.6	$\pm$	0.022	&	(p)	\\
L98-59	c	&	(q)	&	3.6904	$\pm$	0.0003	&	(r)	&	1.35	$\pm$	0.07	&	(r)	&	2.42	$\pm$	0.35	&	(r)	&	5.4	$\pm$	1.2	&	(r)	&	0.312	$\pm$	0.031	&	(r)	&	0.314	$\pm$	0.014	&	(r)	\\
LHS 1140	b	&	(s)	&	24.73694	$\pm$	0.00041	&	(t)	&	1.635	$\pm$	0.046	&	(t)	&	6.38	$\pm$	0.45	&	(t)	&	8.04	$\pm$	0.82	&	(t)	&	0.191	$\pm$	0.012	&	(t)	&	0.2134	$\pm$	0.0035	&	(t)	\\
LHS 1140	c	&	(s)	&	3.777929	$\pm$	0.00003	&	(t)	&	1.169	$\pm$	0.038	&	(t)	&	1.76	$\pm$	0.17	&	(t)	&	6.07	$\pm$	0.78	&	(t)	&	0.191	$\pm$	0.012	&	(t)	&	0.2134	$\pm$	0.0035	&	(t)	\\
LTT 3780	b	&	(u)	&	0.768448	$\pm$	0.000055	&	(u)	&	1.332	$\pm$	0.074	&	(u)	&	2.62	$\pm$	0.47	&	(u)	&	6.1	$\pm$	1.7	&	(u)	&	0.401	$\pm$	0.012	&	(u)	&	0.374	$\pm$	0.011	&	(u)	\\
LTT 3780	c	&	(u)	&	12.2519	$\pm$	0.003	&	(u)	&	2.3	$\pm$	0.16	&	(u)	&	8.6	$\pm$	1.45	&	(u)	&	3.9	$\pm$	1	&	(u)	&	0.401	$\pm$	0.012	&	(u)	&	0.374	$\pm$	0.011	&	(u)	\\
TOI-1235	b	&	(z)	&	3.444729	$\pm$	0.000031	&	(z)	&	1.738	$\pm$	0.083	&	(z)	&	6.91	$\pm$	0.8	&	(z)	&	7.4	$\pm$	1.4	&	(z)	&	0.64	$\pm$	0.016	&	(z)	&	0.63	$\pm$	0.015	&	(z)	\\
TRAPPIST-1	b	&	(w)	&	1.51087081	$\pm$	0.0000006	&	(x)	&	1.121	$\pm$	0.031	&	(y)	&	1.017	$\pm$	0.149	&	(y)	&	0.963	$\pm$	0.122	&	(y)	&	0.089	$\pm$	0.007	&	(y)	&	0.117	$\pm$	0.0036	&	(x)	\\
TRAPPIST-1	c	&	(w)	&	2.4218233	$\pm$	0.0000017	&	(x)	&	1.095	$\pm$	0.03	&	(y)	&	1.156	$\pm$	0.136	&	(y)	&	1.17	$\pm$	0.11	&	(y)	&	0.089	$\pm$	0.007	&	(y)	&	0.117	$\pm$	0.0036	&	(x)	\\
TRAPPIST-1	d	&	(w)	&	4.04961	$\pm$	0.000063	&	(x)	&	0.784	$\pm$	0.023	&	(y)	&	0.297	$\pm$	0.037	&	(y)	&	0.817	$\pm$	0.086	&	(y)	&	0.089	$\pm$	0.007	&	(y)	&	0.117	$\pm$	0.0036	&	(x)	\\
TRAPPIST-1	e	&	(w)	&	6.099615	$\pm$	0.000011	&	(x)	&	0.91	$\pm$	0.026	&	(y)	&	0.772	$\pm$	0.077	&	(y)	&	1.358	$\pm$	0.097	&	(y)	&	0.089	$\pm$	0.007	&	(y)	&	0.117	$\pm$	0.0036	&	(x)	\\
TRAPPIST-1	f	&	(w)	&	9.20669	$\pm$	0.000015	&	(x)	&	1.046	$\pm$	0.029	&	(y)	&	0.934	$\pm$	0.079	&	(y)	&	1.08	$\pm$	0.05	&	(y)	&	0.089	$\pm$	0.007	&	(y)	&	0.117	$\pm$	0.0036	&	(x)	\\
TRAPPIST-1	g	&	(w)	&	12.35294	$\pm$	0.00012	&	(x)	&	1.148	$\pm$	0.032	&	(y)	&	1.148	$\pm$	0.097	&	(y)	&	1.01	$\pm$	0.05	&	(y)	&	0.089	$\pm$	0.007	&	(y)	&	0.117	$\pm$	0.0036	&	(x)	\\
TRAPPIST-1	h	&	(w)	&	18.767	$\pm$	0.004	&	(x)	&	0.773	$\pm$	0.026	&	(y)	&	0.331	$\pm$	0.053	&	(y)	&	0.954	$\pm$	0.15	&	(y)	&	0.089	$\pm$	0.007	&	(y)	&	0.117	$\pm$	0.0036	&	(x)	\\
L231-32	b	&	(z)	&	3.3601538	$\pm$	0.0000048	&	(z)	&	1.206	$\pm$	0.039	&	(z)	&	1.58	$\pm$	0.26	&	(z)	&	4.97	$\pm$	0.94	&	(z)	&	0.386	$\pm$	0.008	&	(z)	&	0.378	$\pm$	0.011	&	(z)	\\
L231-32	c	&	(z)	&	5.6605731	$\pm$	0.0000031	&	(z)	&	2.355	$\pm$	0.064	&	(z)	&	6.15	$\pm$	0.37	&	(z)	&	2.60	$\pm$	0.26	&	(z)	&	0.386	$\pm$	0.008	&	(z)	&	0.378	$\pm$	0.011	&	(z)	\\
L231-32	d	&	(z)	&	11.379573	$\pm$	0.000013	&	(z)	&	2.133	$\pm$	0.058	&	(z)	&	4.78	$\pm$	0.43	&	(z)	&	2.72	$\pm$	0.33	&	(z)	&	0.386	$\pm$	0.008	&	(z)	&	0.378	$\pm$	0.011	&	(z)	\\

\hline
\end{tabular}
\end{table*}

\newpage

\label{lastpage}
\end{document}